\documentclass{article}
\usepackage{graphicx} 
\usepackage{subcaption}
\usepackage{bm}
\usepackage[margin=25truemm]{geometry}
\usepackage{amsmath}
\usepackage{mathtools}
\usepackage{mathrsfs}
\usepackage{braket}
\usepackage{tikz}
\usepackage{float}
\usepackage{physics}
\usepackage{authblk}

\usepackage{hyperref}
\usepackage[title]{appendix}

\newcommand{\defeq}{\overset{\text{def}}{=}}

\usetikzlibrary{intersections,calc,arrows.meta}
\numberwithin{equation}{section}
\title{Consistency relations of amplitude and phase fluctuations of gravitational waves magnified by strong gravitational lensing}
\author{Yuta Nakazono$^{1}$, Teruaki Suyama$^{1}$}
\affil{$^{1}$Department of Physics, Institute of Science Tokyo, 2-12-1 Ookayama, Meguro-ku, Tokyo 152-8551, Japan}
\date{}
\begin{document}
\maketitle
\vspace{10pt}

\begin{abstract}
We discuss the amplitude and phase fluctuations of gravitational waves due to wave optics lensing in the presence of both a strong lens and cosmological weak lenses. By applying the geometric optics approximation to the strong lens and treating the weak lensing potential perturbatively, we obtain the amplification factor up to the second order in the weak lensing potential. Additionally, we establish a methodology to systematically evaluate the weak lensing effects based on diagrammatic rules. Based on the derived amplification factor, we evaluate the statistics of the fluctuations and demonstrate that the consistency relations originally established in the absence of a strong lens still hold in exactly the same form when a strong lens is present. The physical origin of these relations is also discussed. Furthermore, we demonstrate that for the mean of the weak lensing signal, both the magnification of the signal and the shift of the Fresnel scale to larger scales occur, consistent with the behavior observed in the variance.
\end{abstract}

\section{INTRODUCTION}
The first direct detection of gravitational waves in 2015 by LIGO \cite{LIGO} marked the dawn of a new era: gravitational wave astronomy. Just as with electromagnetic waves, the propagation of gravitational waves is affected by the gravitational lensing effect, caused by intervening matter between the source and the observer. However, since the wavelengths of gravitational waves can be significantly longer than those of electromagnetic waves, wave optics effects such as diffraction and interference can become crucial depending on the lens mass and the wave frequency (see e.g., \cite{Nakamura,oguri:wave} for reviews). These effects are quantitatively described by the amplification factor, which modifies the amplitude and phase of the wave. 
Moreover, it was demonstrated in \cite{takahashi:wave,oguri:probe} that this fluctuation is significantly affected by a characteristic scale known as the Fresnel scale, and the potential for probing dark matter fluctuations at the sub-galactic scale was discussed. In addition, recent research has explored the possibility of probing scales smaller than the Fresnel scale by utilizing specific quantities derived from the amplification factor \cite{tanaka}. While the Cold Dark Matter (CDM) model has been remarkably successful in explaining the large-scale structure of the universe, the nature of structures at sub-galactic scales remains a subject of ongoing debate. Probing the sub-galactic scale through the gravitational lensing of gravitational waves offers a unique opportunity to explore this regime, which is largely inaccessible to conventional electromagnetic observations. Such a new observational window is expected to provide critical insights into the nature of dark matter.

Motivated by these observational prospects, theoretical studies have also been making steady progress. Pioneering theoretical work by \cite{takahashi:wave} investigated the weak lensing of gravitational waves as they propagate through inhomogeneous cosmological matter fluctuations. Using the Born approximation, which is a first-order approximation, the author derived a relation between the matter power spectrum and the variance of amplitude and phase fluctuations. Building upon this, a previous study \cite{mizuno:beyond} employed the post-Born approximation to show that while the means of amplitude and phase fluctuations vanish at first order, they become non-zero at second order, where relationships with the matter power spectrum are established. Furthermore, universal consistency relations between the variances and the means of amplitude and phase fluctuations were established \cite{inamori,mizuno:new}, which hold independently of the specific form of the matter power spectrum. The physical origin of these relations was also clarified in \cite{mizuno:new} to be energy conservation and the Shapiro time delay. These consistency relations are of significant practical importance as they serve as a robust diagnostic tool for signal verification. Specifically, they provide a criterion to distinguish genuine gravitational wave lensing signals, such as amplitude and phase fluctuations, from non-physical artifacts, such as instrumental noise or systematic errors. While physical lensing signals must satisfy these relations, any violation of the consistency relations for observed signals would indicate that the detected fluctuations are false. Consequently, these relations offer a critical self-consistency test for validating future gravitational wave lensing observations. 

Separately, the work in \cite{oguri:strong} derived the amplification factor in the presence of a strong lens by applying the Born approximation to the weak lensing from subhalos, and subsequently showed that the presence of the strong lens enhances the amplitude and phase shifts caused by these weak lensing effects. This effect is expected to enhance the observability of the weak lensing signal. 

As discussed above, strong lensing enhances the observability of amplitude and phase fluctuations induced by weak lensing. Since these intrinsic fluctuations are typically too weak to be detected in isolation, utilizing the magnification provided by a strong lens represents a promising strategy to overcome the detection threshold. Consequently, future observations are expected to capture such magnified signals. To employ consistency relations as a diagnostic tool for these observations, it is essential to verify whether they remain valid in the presence of strong lensing.

\begin{figure}[t!]
    \centering 
    \includegraphics[width=\textwidth]{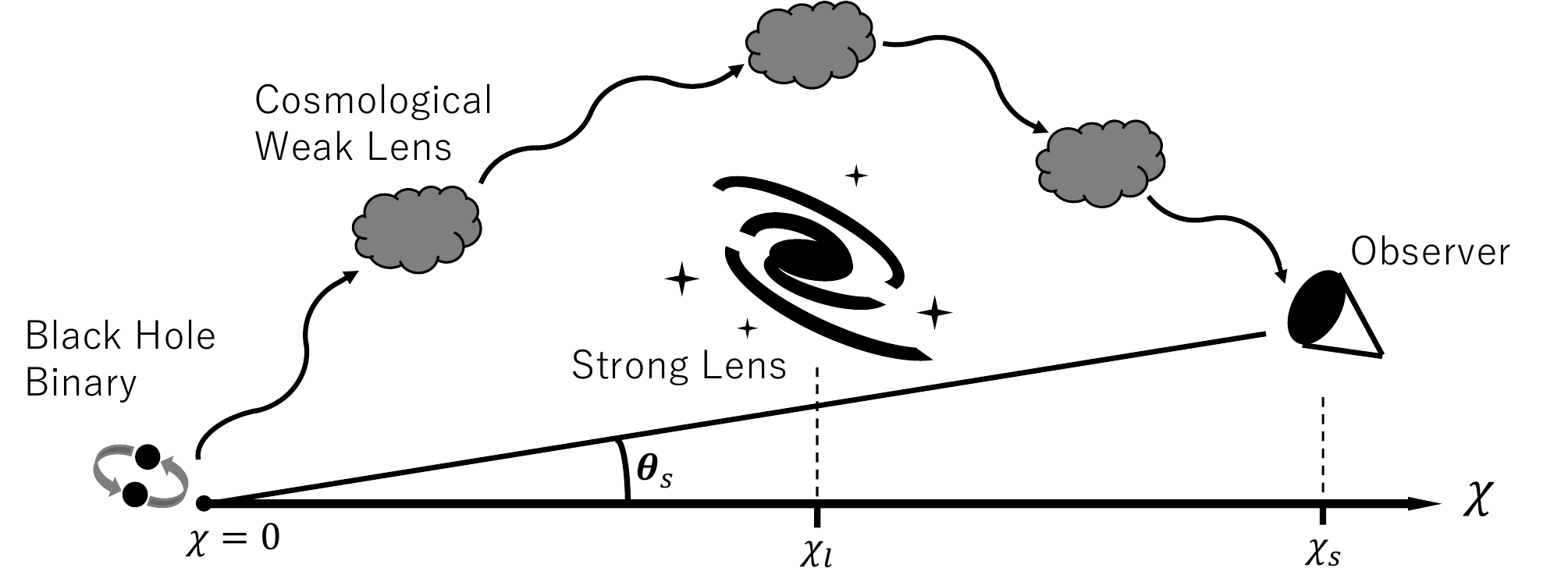}
    \caption{Gravitational lensing geometry. We adopt a coordinate system $(\chi,\bm{\theta})$ with the origin at the source, where $\chi$ is the comoving distance from the source and $\bm{\theta}$ is the two-dimensional vector perpendicular to the line of sight. The observer's position is denoted by $(\chi_{s},\bm{\theta}_{s})$. The strong lens component is treated using the thin-lens and geometric optics approximations, while the weak lens potential is treated perturbatively within the framework of wave optics.}
    \label{fig:gwimage}
\end{figure}

In this paper, we analytically derive the amplification factor for the situation where both strong and weak lensing are present as shown in Fig. \ref{fig:gwimage}. We employ the thin-lens approximation and the stationary phase approximation for the strong lens component and treat the weak lens potential as a small perturbation relative to the strong lens potential. While the previous study \cite{oguri:strong} focused on weak lensing effects from subhalos surrounding a strong lens, we consider the more general case of cosmological density fluctuations. Furthermore, whereas that work \cite{oguri:strong} derived the amplification factor only up to the first order in weak lensing to obtain the variance, we expand the amplification factor to the second order. This allows us to calculate not only the variance but also the mean, which is essential for a verification of the consistency relations. From this amplification factor, we derive formulae for key lensing statistics, namely the mean and variance of the amplitude and phase fluctuations, and demonstrate that our derived formulae satisfy the consistency relations of the same form as those in previous works \cite{inamori,mizuno:new}. We also show that the physical origin of the consistency relations is the Shapiro time delay and the law of energy conservation. In addition, we investigate the dependence of these lensing statistics on the strong lensing magnification factor. Throughout this study, we adopt the units in which $c=1$, and assume a flat Universe with the dark matter density $\Omega_{c0}h^{2}=0.125$, the baryon density $\Omega_{b0}h^{2}=0.022$, and the dimensionless Hubble constant $h=0.7$.

\section{Formulation}

In this section, we obtain the explicit form of the amplification factor for the case where there are a strong lens such as galaxies and weak lenses due to cosmological density fluctuations, as illustrated in Fig. \ref{fig:gwimage}. We begin in Subsection 2.1 by reviewing the general theory of the amplification factor \cite{Nakamura,oguri:wave} and discussing the representation of the amplification factor using unitary operators based on the similarity between the equation followed by the amplification factor and the Schr\"{o}dinger equation. In Subsection 2.2, we state what kind of approximations are used for strong and weak lenses respectively, and also discuss how to separate the unitary operators corresponding to them. Subsequently, in Subsection 2.3, we describe a method for perturbatively treating weak lensing by analogy with the interaction representation in quantum mechanics, and obtain an expression for the amplification factor expanded to the first order of the weak lensing potential. Finally, in Subsection 2.4, we discuss that the calculation of higher-order terms for weak lensing can be obtained systematically by using a diagrammatic method similar to Feynman diagrams in quantum field theory. We obtain an expression for the amplification factor expanded to the second order of the weak lens using the diagrams.

\subsection{Amplification factor and unitary operator}

Assuming a monochromatic wave with the comoving angular frequency $\omega$ given by $\phi(t,\bm{x})=\Tilde{\phi}(\bm{x})e^{-i\omega \eta}$, where $\eta$ is the conformal time, the amplification factor $F(\omega,\bm{x})$, which describes the effects of gravitational lensing, is defined as follows \cite{Nakamura}:
\begin{align}
    F(\omega,\bm{x})\defeq\frac{\Tilde{\phi}}{\Tilde{\phi_{0}}},
\end{align}
where $\Tilde{\phi}_{0}$ denotes the original waveform in the absence of gravitational lensing and $\Tilde{\phi}$ is the observed waveform including gravitational lensing effects. As established in previous study \cite{Nakamura}, by adopting spherical coordinates $(\chi,\bm{\theta})$ centered at the source, the amplification factor is found to satisfy the following propagation equation:
\begin{align}
    i\frac{1}{\omega}\frac{\partial }{\partial\chi}F(\chi,\bm{\theta})=\qty(-\frac{1}{2\chi^{2}\omega^{2}}\Delta_{\bm{\theta}}+2\Phi(\chi,\bm{\theta}))F(\chi,\bm{\theta}).\label{Feq}
\end{align}
Here, $\chi$ is the comoving radial distance, $\Phi(\chi,\bm{\theta})$ is the gravitational potential, and $\Delta_{\bm{\theta}}$ represents the two-dimensional Laplacian with respect to the angular coordinates $\bm{\theta}$.
Equation \eqref{Feq} is analogous to the two-dimensional Schrödinger equation with a time-dependent potential, which is generally expressed as:
\begin{align}
    i\hbar\frac{\partial}{\partial t}\Psi(t,\bm{x})=\qty(-\frac{\hbar^{2}}{2m}\Delta_{\bm{x}}+V(t,\bm{x}))\Psi(t,\bm{x}),
\end{align}
where $t, \bm{x}, \hbar, m, V,$ and $\Psi$ denote time, two-dimensional position vector, reduced Planck constant, mass, potential, and wave function, respectively. By introducing the following correspondences:
\begin{align}
t\leftrightarrow \chi,\quad \bm{x}\leftrightarrow \bm{\theta},\quad \hbar\leftrightarrow\frac{1}{\omega},\quad  m\leftrightarrow\chi^{2},\quad  V\leftrightarrow2\Phi,\quad \Psi\leftrightarrow F,
\end{align}
Equation \eqref{Feq} can be regarded as a Schrödinger equation in two-dimensional space with a time-dependent mass $m=\chi^2$ and potential $2\Phi(\chi,\bm{\theta})$. Here, using the coordinate basis $\ket{\bm{\theta}}$, we consider a state $\ket{F(\chi)}$ satisfying $F(\chi,\bm{\theta})=\bra{\bm{\theta}}\ket{F(\chi)}$. Writing Eq. \eqref{Feq} without the coordinate basis yields the following equation for $\ket{F(\chi)}$:
\begin{align}
    i\frac{1}{\omega}\frac{d }{d\chi}\ket{F(\chi)}=\qty(\frac{\hat{\bm{p}}^{2}}{2\chi^{2}}+2\hat{\Phi}(\chi,\hat{\bm{\theta}}))\ket{F(\chi)}.\label{Feqket}
\end{align}
Here, $\hat{\bm{\theta}}, \hat{\bm{p}},$ and $\hat{\Phi}(\chi,\hat{\bm{\theta}})$ denote the position, momentum, and gravitational potential operators, respectively. In the coordinate representation, these operators correspond to $\bm{\theta}, \frac{1}{i\omega}\bm{\nabla}_{\bm{\theta}},$  and $\Phi(\chi,\bm{\theta}$).
Consequently, in analogy with standard quantum mechanics, the solution to Eq. \eqref{Feqket} can be expressed as $\ket{F(\chi)}=\hat{U}(\chi,\chi_{0})\ket{F(\chi_{0})}$ using the unitary operator:
\begin{align}
    \hat{U}(\chi,\chi_{0})=T\exp{-i\omega\int^{\chi}_{\chi_{0}}d\chi'\qty(\frac{\hat{\bm{p}}^{2}}{2\chi^{'2}}+2\hat{\Phi}(\chi',\hat{\bm{\theta}}))},\label{uni}
\end{align}
where $T$ denotes the time-ordering operator. The amplification factor $F(\chi,\bm{\theta})$ corresponds to the projection of this solution $\ket{F(\chi)}$ onto the basis of angular coordinates $\bra{\bm{\theta}}$. Considering a wave emitted from the source at $\chi_{0}=0$, and setting the observer's position at $(\chi_{s},\bm{\theta}_{s})$ as shown in Figure \ref{fig:gwimage}, the amplification factor is expressed as follows:
\begin{align}
    F(\chi_{s},\bm{\theta}_{s})=\bra{\bm{\theta}_{s}}\hat{U}(\chi_{s},0)\ket{F(\chi_{0}=0)}.\label{Fsol1}
\end{align}
Note that the amplification factor $F$ is a function of $\omega$, though this dependence is omitted for brevity.
Since the scalar wave is unaffected by gravitational lensing at the moment of emission from the source, we impose the initial condition $F(\chi=0,\bm{\theta})=1$. By inserting the completeness relation $\int d\bm{\theta}_{0}\ket{\bm{\theta}_{0}}\bra{\bm{\theta}_{0}}=1$ into Eq. \eqref{Fsol1}, we obtain the following expression:
\begin{align}
    F(\chi_{s},\bm{\theta}_{s})=\int d\bm{\theta}_{0}\bra{\bm{\theta}_{s}}\hat{U}(\chi_{s},0)\ket{\bm{\theta}_{0}}.\label{Fsol2}
\end{align}

\subsection{Amplification factor in the presence of strong and weak gravitational lensing}
Based on the general formulation presented in Subsection 2.1, we now consider the amplification factor for a scalar wave propagating through the gravitational fields of both a strong lens, such as a galaxy or galaxy cluster, and weak lensing from large-scale structure as shown in Figure \ref{fig:gwimage}. We first distinguish between the strong lensing potential caused by a local massive object such as a galaxy and the cosmological weak lensing potential. For the strong lens component, we employ the thin-lens and stationary phase approximations, treating it within the framework of geometric optics. In contrast, the weak lens potential is assumed to have a relatively small effect on the amplification factor and is therefore treated perturbatively within the framework of wave optics. Following this approach, we decompose the total gravitational potential into a strong lensing potential $\psi$ and a weak lensing gravitational potential $\Phi_{W}$ as follows:
\begin{align}
    \hat{\Phi}(\chi,\hat{\bm{\theta}})=\frac{1}{2}\hat{\psi}(\hat{\bm{\theta}})\delta(\chi-\chi_{l})+\hat{\Phi}_{W}(\chi,\hat{\bm{\theta}}).\label{grapo}
\end{align}
Here, $\chi_{l}$ denotes the comoving radial distance from the source to the strong lens.  
Corresponding to this decomposition of the gravitational potential, we also divide the unitary operator into components for the strong and weak lenses. Firstly, we decompose the unitary operator for the entire interval into three parts to isolate the strong lens at $\chi_l$, using an infinitesimal interval $\Delta\chi$ as follows:
\begin{align}
    \hat{U}(\chi_{s},0)=\hat{U}(\chi_{s},\chi_{l}+\Delta \chi)\hat{U}(\chi_{l}+\Delta \chi,\chi_{l}-\Delta \chi)\hat{U}(\chi_{l}-\Delta \chi,0).\label{unil}
\end{align}
By substituting Eqs. \eqref{uni} and \eqref{grapo} into Eq. \eqref{unil} and taking the limit $\Delta\chi\rightarrow0$, the unitary operator at $\chi_{l}$ can be evaluated as:
\begin{align}
    \hat{U}(\chi_{l}+\Delta \chi,\chi_{l}-\Delta \chi)&=T\exp{-i\omega\int^{\chi_{l}+\Delta\chi}_{\chi_{l}-\Delta\chi}d\chi\qty(\frac{\hat{p}^{2}}{2\chi^{2}}+\hat{\psi}(\hat{\bm{\theta}})\delta(\chi-\chi_{l})+2\hat{\Phi}_{W}(\chi,\hat{\bm{\theta}}))}\nonumber\\
    &\rightarrow\exp{-i\omega\hat{\psi}(\hat{\bm{\theta}})}.
\end{align}
Substituting these into Eq. \eqref{Fsol2} and inserting the completeness relations at each boundary, the amplification factor is expressed in the following integral form:
\begin{align}
    F(\chi_{s},\bm{\theta}_{s})&=\int d\bm{\theta}_{0}\bra{\bm{\theta}_{s}}\hat{U}(\chi_{s},\chi_{l})\qty(\int d\bm{\theta}_{a}\ket{\bm{\theta}_{a}}\bra{\bm{\theta}_{a}})\exp{-i\omega\hat{\psi}(\hat{\bm{\theta}})}\qty(\int d\bm{\theta}_{b}\ket{\bm{\theta}_{b}}\bra{\bm{\theta}_{b}})\hat{U}(\chi_{l},0)\ket{\bm{\theta}_{0}}\nonumber\\
    &=\int d\bm{\theta}_{0}d\bm{\theta}_{l}\bra{\bm{\theta}_{s}}\hat{U}(\chi_{s},\chi_{l})\ket{\bm{\theta}_{l}}\exp{-i\omega\psi(\bm{\theta}_{l})}\bra{\bm{\theta}_{l}}\hat{U}(\chi_{l},0)\ket{\bm{\theta}_{0}}.\label{F}
\end{align}
Here, we have replaced $\bm{\theta}_{a}$ with $\bm{\theta}_{l}$ to clarify that the integration is performed over the lens plane where the strong lens is situated. Since the strong lensing potential exists only at $\chi_{l}$, the propagators, such as $\bra{\bm{\theta}_{s}}\hat{U}(\chi_{s},\chi_{l})\ket{\bm{\theta}_{l}}$, depend only on the weak lensing potential along the propagation path.

\subsection{Interaction Picture and Perturbative Expansion}
In this subsection, we derive an explicit expression for the propagator by treating the weak lensing potential perturbatively. To improve the clarity of the discussion, we employ an analogy with the interaction picture in quantum mechanics. In the Schrödinger equation \eqref{Feqket}, we decompose the Hamiltonian $\hat{H}$ acting on $F$ into a free-particle part $\hat{H}_{0}$ and an interaction part $\hat{V}$ as $\hat{H}=\hat{H}_{0}+\hat{V}$. Accordingly, the unitary operator can be divided into a free-particle part $\hat{U}_{0}$ and an interaction part $\hat{U}_{\text{int}}$ such that $\hat{U}(\chi,\chi_{0})=\hat{U}_{0}(\chi,\chi_{0})\hat{U}_{\text{int}}(\chi,\chi_{0})$. Since $\hat{U}_{0}$ satisfies the free-particle Schrödinger equation:
\begin{align}
    i\frac{1}{\omega}\frac{d}{d\chi}\hat{U}_{0}(\chi,\chi_{0})=\frac{\hat{p}^{2}}{2\chi^{2}}\hat{U}_{0}(\chi,\chi_{0}),\label{unieqfree}
\end{align}
the interaction part $\hat{U}_{\text{int}}$ obeys the following equation:
\begin{align}
    i\frac{1}{\omega}\frac{d}{d\chi}\hat{U}_{\text{int}}(\chi,\chi_{0})=\hat{V}_{\text{int}}(\chi,\chi_{0})\hat{U}_{\text{int}}(\chi,\chi_{0}),\label{unieqint}
\end{align}
where $\hat{V}_{\text{int}}(\chi,\chi_{0})=\hat{U}_{0}^{-1}(\chi,\chi_{0})\hat{V}(\chi)\hat{U}_{0}(\chi,\chi_{0})$. The solution to Eq. \eqref{unieqint} is given by 
\begin{align}
    \hat{U}_{\text{int}}(\chi,\chi_{0})=T\exp{-i\omega\int^{\chi}_{\chi_{0}}d\chi'\hat{V}_{\text{int}}(\chi',\chi_{0})}.\label{uniint}
\end{align}
As stated previously, the propagator depends only on the weak lensing potential; thus, we set $\hat{V}=2\hat{\Phi}_{W}$. Accordingly, by performing a Taylor expansion of Eq. \eqref{uniint} with respect to $\hat{\Phi}_{W}$,  the expansion of the unitary operator $\hat{U}$ is obtained as follows:
\begin{align}
    \hat{U}(\chi,\chi_{0})&=\hat{U}_{0}(\chi,\chi_{0})\left\{1-2i\omega\int^{\chi}_{\chi_{0}}d\chi_{a} \hat{\Phi}_{W,\text{int}}(\chi_{a},\chi_{0},\hat{\bm{\theta}})\right.\nonumber\\
    &\left.+\frac{(-2i\omega)^{2}}{2}\int^{\chi}_{\chi_{0}}d\chi_{a}\int^{\chi}_{\chi_{0}}d\chi_{b}T\qty[\hat{\Phi}_{W,\text{int}}(\chi_{a},\chi_{0},\hat{\bm{\theta}})\hat{\Phi}_{W,\text{int}}(\chi_{b},\chi_{0},\hat{\bm{\theta}})]\right\}+\mathcal{O}(\hat{\Phi}_{W,\text{int}}^{3}).\label{U}
\end{align}
Here, $\hat{\Phi}_{W,\text{int}}(\chi,\chi_{0},\hat{\bm{\theta}})=\hat{U}_{0}^{-1}(\chi,\chi_{0})\hat{\Phi}_{W}(\chi,\hat{\bm{\theta}})\hat{U}_{0}(\chi,\chi_{0})$.

Before proceeding to the specific perturbative calculations, we derive the free-particle propagator. The solution to Eq. \eqref{unieqfree} can be expressed as:
\begin{align}
    \hat{U}_{0}(\chi_{b},\chi_{a})=T\exp{-i\omega\int^{\chi_{b}}_{\chi_{a}}d\chi\frac{\hat{p}^{2}}{2\chi^{2}}}=\exp{-i\omega\frac{\chi_{b}-\chi_{a}}{2\chi_{a}\chi_{b}}\hat{p}^{2}},
\end{align}
where $\chi_{a}$ and $\chi_{b}$ are arbitrary radial distances.
Using this expression, the free-particle propagator in the coordinate representation, $\bra{\bm{\theta}_{b}}\hat{U}_{0}(\chi_{b},\chi_{a})\ket{\bm{\theta}_{a}}$ is obtained as:
\begin{align}
    \bra{\bm{\theta}_{b}}\hat{U}_{0}(\chi_{b},\chi_{a})\ket{\bm{\theta}_{a}}=\frac{\omega}{2\pi i}\frac{\chi_{a}\chi_{b}}{\chi_{b}-\chi_{a}}\exp{\frac{i\omega}{2}\frac{\chi_{a}\chi_{b}}{\chi_{b}-\chi_{a}}|\bm{\theta}_{a}-\bm{\theta}_{b}|^{2}}.\label{propag}
\end{align}
In the above derivation, we have utilized the completeness relation for the momentum eigenstates, $\int d\bm{p}\ket{\bm{p}}\bra{\bm{p}}=1$, and the free-particle wave function, $\bra{\bm{\theta}}\ket{\bm{p}}=\frac{\omega}{2\pi}\exp{i\omega\bm{p}\cdot\bm{\theta}}$.

With these preparations, we now perform a specific evaluation of the amplification factor. Firstly, we determine the amplification factor in the absence of the weak lensing potential, denoted as $F_{0}$. By replacing the unitary operator in Eq. \eqref{F} with the zeroth-order term of $\Phi_{W,\text{int}}$ in Eq. \eqref{U}---namely, the free-particle operator $\hat{U}_{0}$---and employing the propagator from Eq. \eqref{propag}, we obtain $F_{0}$ in the following form:
\begin{align}
    F_{0}(\chi_{s},\bm{\theta}_{s})&=\int d\bm{\theta}_{\epsilon}d\bm{\theta}_{l}\bra{\bm{\theta}_{s}}\hat{U}_{0}(\chi_{s},\chi_{l})\ket{\bm{\theta}_{l}}\exp{-i\omega\psi(\bm{\theta}_{l})}\bra{\bm{\theta}_{l}}\hat{U}_{0}(\chi_{l},\epsilon)\ket{\bm{\theta}_{\epsilon}}\nonumber\\
    &=\frac{\omega}{2\pi i}\frac{\chi_{l}\chi_{s}}{\chi_{s}-\chi_{l}}\int d\bm{\theta}_{l}\exp{i\omega\qty[\frac{\chi_{l}\chi_{s}}{2(\chi_{s}-\chi_{l})}|\bm{\theta}_{l}-\bm{\theta}_{s}|^{2}-\psi(\bm{\theta}_{l})]}.\label{F0}
\end{align}
To avoid the singularity of the propagator at the origin $\chi=0$, we have regularized the expression by setting the source's position at $\chi=\epsilon>0$, where $\epsilon$ is an infinitesimal quantity. Accordingly, we have replaced $\bm{\theta}_{0}$ with $\bm{\theta}_{\epsilon}$. The limit $\epsilon\rightarrow0$ is then taken after evaluating the integral in the first line. This expression for $F_{0}$ is consistent with the amplification factor obtained in previous study \cite{Nakamura} for strong lensing systems where the thin-lens approximation is applicable.
Regarding the strong gravitational lens, following previous study \cite{oguri:strong}, we consider a small region around the $j$-th multiple image at $\bm{\theta}_{l,j}$ for a strong lens system. We use the following relation to apply the stationary phase approximation:
\begin{align}
    \Delta t_{S}(\bm{\theta}_{l})\defeq\frac{\chi_{l}\chi_{s}}{2(\chi_{s}-\chi_{l})}|\bm{\theta}_{l}-\bm{\theta}_{s}|^{2}-\psi(\bm{\theta}_{l})\simeq\Delta t_{S}(\bm{\theta}_{l,j})+\frac{\chi_{l}\chi_{s}}{2(\chi_{s}-\chi_{l})}(\mu_{j,1}^{-1}\alpha_{1}^{2}+\mu_{j,2}^{-1}\alpha_{2}^{2}), \label{spa}
\end{align}
where $\bm{\theta}_{l,j}$ is the $j$-th image position, $\mu_{j,1},\mu_{j,2}$ are the magnification factors of the $j$-th image among the multiple images produced by the strong lens, $\Delta t_{S}$ is the Shapiro time delay due to the strong lens, and $\bm{\theta}_{l}=\bm{\theta}_{l,j}+\bm{\alpha}$. Using Eq. \eqref{spa}, the amplification factor for the $j$-th image is derived from Eq. \eqref{F0} as follows:
\begin{align}
    F_{j0}(\chi_{s},\bm{\theta}_{s})=|\mu_{j}|^{1/2}e^{i\omega\Delta t_{S}(\bm{\theta}_{l,j})-i\pi n_{j}\text{sgn}(\omega)},\label{F0j}
\end{align}
where $\mu_{j}\defeq \mu_{j,1}\mu_{j,2}$, $n_{j}=0,1/2,1$ corresponds to minimum, saddle point, and maximum images, respectively, and the $\text{sgn}(\omega)$ is the sign function of $\omega$.

Next, we derive the first-order correction term $F_{1}$ due to the weak lensing potential. This is derived by substituting the first-order term with respect to $\Phi_{W,\text{int}}$ in Eq. \eqref{U} for either of the two unitary operators appearing in Eq. \eqref{F}. By summing over all such possible combinations, the first-order correction $F_{1}$ is obtained as follows:
\begin{align}
    F_{1}(\chi_{s},\bm{\theta}_{s})&=\int d\bm{\theta}_{\epsilon}d\bm{\theta}_{l}\bra{\bm{\theta}_{s}}\hat{U}_{0}(\chi_{s},\chi_{l})\ket{\bm{\theta}_{l}}\exp{-i\omega\psi(\bm{\theta}_{l})}\bra{\bm{\theta}_{l}}\hat{U}_{0}(\chi_{l},\epsilon)\frac{2\omega}{i}\int^{\chi_{l}}_{\epsilon}d\chi_{a}\hat{\Phi}_{W,\text{int}}(\chi_{a},\epsilon,\hat{\bm{\theta}})\ket{\bm{\theta}_{\epsilon}}\nonumber\\
    &+\int d\bm{\theta}_{\epsilon}d\bm{\theta}_{l}\bra{\bm{\theta}_{s}}\hat{U}_{0}(\chi_{s},\chi_{l})\frac{2\omega}{i}\int^{\chi_{s}}_{\chi_{l}}d\chi_{a}\hat{\Phi}_{W,\text{int}}(\chi_{a},\chi_{l},\hat{\bm{\theta}})\ket{\bm{\theta}_{l}}\exp{-i\omega\psi(\bm{\theta}_{l})}\bra{\bm{\theta}_{l}}\hat{U}_{0}(\chi_{l},\epsilon)\ket{\bm{\theta}_{\epsilon}}.\label{F10}
\end{align}
To simplify Eq. \eqref{F10}, we rewrite the propagator in the first term, which describes the propagation from the source to the strong lens $(\chi_{a}:\epsilon\rightarrow\chi_{l})$, as follows:
\begin{align}
    &\bra{\bm{\theta}_{l}}\hat{U}_{0}(\chi_{l},\epsilon)\frac{2\omega}{i}\int^{\chi_{l}}_{\epsilon}d\chi_{a}\hat{U}_{0}^{-1}(\chi_{a},\epsilon)\hat{\Phi}_{W}(\chi_{a},\hat{\bm{\theta}})\hat{U}_{0}(\chi_{a},\epsilon)\ket{\bm{\theta}_{\epsilon}}\nonumber\\
    &=\int^{\chi_{l}}_{\epsilon}d\chi_{a}\int d\bm{\theta}_{a}\bra{\bm{\theta}_{l}}\hat{U}_{0}(\chi_{l},\chi_{a})\ket{\bm{\theta}_{a}}\frac{2\omega}{i}\Phi_{W}(\chi_{a},\bm{\theta}_{a})\bra{\bm{\theta}_{a}}\hat{U}_{0}(\chi_{a},\epsilon)\ket{\bm{\theta}_{\epsilon}}.
\end{align}
In this transformation, we have inserted the completeness relation for the coordinate representation. Similarly, the second term in Eq. \eqref{F10}, which describes the propagation from the strong lens to the observer $(\chi_{a}:\chi_{l}\rightarrow\chi_{s})$, can be rewritten as follows:
\begin{align}
    &\bra{\bm{\theta}_{s}}\hat{U}_{0}(\chi_{s},\chi_{l})\frac{2\omega}{i}\int^{\chi_{s}}_{\chi_{l}}d\chi_{a}\hat{U}_{0}^{-1}(\chi_{a},\chi_{l})\hat{\Phi}_{W}(\chi_{a},\hat{\bm{\theta}})\hat{U}_{0}(\chi_{a},\chi_{l})\ket{\bm{\theta}_{l}}\nonumber\\
    &=\int^{\chi_{s}}_{\chi_{l}}d\chi_{a}\int d\bm{\theta}_{a}\bra{\bm{\theta}_{s}}\hat{U}_{0}(\chi_{s},\chi_{a})\ket{\bm{\theta}_{a}}\frac{2\omega}{i}\Phi_{W}(\chi_{a},\bm{\theta}_{a})\bra{\bm{\theta}_{a}}\hat{U}_{0}(\chi_{a},\chi_{l})\ket{\bm{\theta}_{l}}.
\end{align}
By utilizing these results and the free-particle propagator from Eq. \eqref{propag}, we can derive the explicit expression for $F_{1}$ as follows:
\begin{align}
    F_{1}(\chi_{s},\bm{\theta}_{s})&=\int^{\chi_{l}}_{\epsilon} d\chi_{a}\int d\bm{\theta}_{l}\int d\bm{\theta}_{a}\int d\bm{\theta}_{\epsilon}\bra{\bm{\theta}_{s}}\hat{U}_{0}(\chi_{s},\chi_{l})\ket{\bm{\theta}_{l}}\exp{-i\omega\psi(\bm{\theta}_{l})}\bra{\bm{\theta}_{l}}\hat{U}_{0}(\chi_{l},\chi_{a})\ket{\bm{\theta}_{a}}\nonumber\\
    &\times\frac{2\omega}{i}\Phi_{W}(\chi_{a},\bm{\theta}_{a})\bra{\bm{\theta}_{a}}\hat{U}_{0}(\chi_{a},\epsilon)\ket{\bm{\theta}_{\epsilon}}\nonumber\\
    &+\int^{\chi_{s}}_{\chi_{l}} d\chi_{a}\int d\bm{\theta}_{l}\int d\bm{\theta}_{a}\int d\bm{\theta}_{\epsilon}\bra{\bm{\theta}_{s}}\hat{U}_{0}(\chi_{s},\chi_{a})\ket{\bm{\theta}_{a}}\frac{2\omega}{i}\Phi_{W}(\chi_{a},\bm{\theta}_{a})\bra{\bm{\theta}_{a}}\hat{U}_{0}(\chi_{a},\chi_{l})\ket{\bm{\theta}_{l}}\nonumber\\
    &\times\exp{-i\omega\psi(\bm{\theta}_{l})}\bra{\bm{\theta}_{l}}\hat{U}_{0}(\chi_{l},\epsilon)\ket{\bm{\theta}_{\epsilon}}.\label{F1}
\end{align}

\subsection{Diagrammatic Formalism}
\begin{figure}[t!]
    \centering 
    \includegraphics[width=0.8\textwidth]{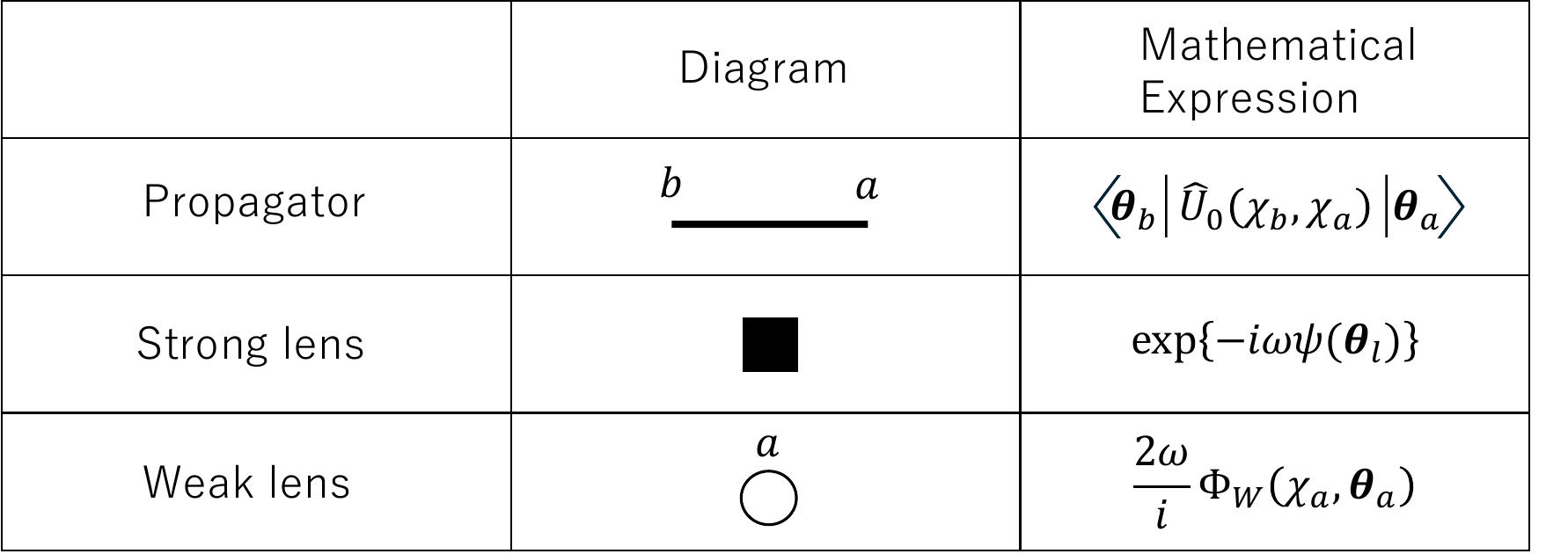}
    \caption{Summary of diagrammatic rules and their corresponding mathematical expressions. The table defines the propagator and the two distinct types of vertices representing the strong and weak lensing potentials, respectively. The explicit mathematical form of the propagator is given by Eq. \eqref{propag}. The amplification factor is constructed by arranging these diagrammatic elements.}
    \label{fig:diag1}
\end{figure}

With the aim of extending our analysis to second-order and higher, we construct a diagrammatic representation based on Eqs. \eqref{F0} and \eqref{F1}. Prior to the diagrammatic construction, we clarify the physical interpretation of the two terms in Eq. \eqref{F1}. In the first term, the comoving radial distance $\chi_{a}$ of the weak lensing potential ranges from $\epsilon$ to $\chi_{l}$. This term describes a process in which the wave is first scattered by the weak lensing potential along the path and subsequently passes through the strong lens. In contrast, the second term describes the reverse sequence: the wave first interacts with the strong lens at $\chi_{l}$ and subsequently experiences the effects of the weak lensing potential in the region between the lens plane and the observer. To clarify these two distinct physical scenarios and to systematically simplify perturbative calculations, we introduce diagrams corresponding to the propagator and vertices representing the gravitational potential, the rules of which are summarized in Figure \ref{fig:diag1}.
Furthermore, to construct the amplification factor from these diagrams, we establish the following set of rules: 

1. In all diagrams, the wave is assumed to propagate from right to left. Specifically, the right and left ends correspond to $(\epsilon,\bm{\theta}_{\epsilon})$ and $(\chi_{s},\bm{\theta}_{s})$, respectively, while the strong lens vertex is located at $(\chi_{l},\bm{\theta}_{l})$.

2. All angular variables $\bm{\theta}$, except for the observer's position $\bm{\theta}_{s}$, are to be integrated.

3. The weak lensing variables $\chi$ are integrated over their respective allowed ranges: from $\epsilon$ to $\chi_{l}$ if the weak lens is located to the right of the strong lens, and from $\chi_{l}$ to $\chi_{s}$ if it is to the left. For second-order or higher terms, where two or more weak lensing vertices appear, we assign dummy variables such as $\chi_{a}$ and $\chi_{b}$ to these positions and perform the integration while accounting for their relative ordering along the line of sight. For instance, in the second-order terms, if both vertices are located between the source and the strong lens $(\epsilon<\chi_{a}<\chi_{b}<\chi_{l})$, the integration is performed as follows:
\begin{align}
    \int^{\chi_{l}}_{\epsilon}d\chi_{b}\int^{\chi_{b}}_{\epsilon}d\chi_{a}\cdots.
\end{align}
By combining these rules with the diagrams above, the explicit expression for the amplification factor can be systematically constructed.

Using these diagrams and rules, $F_{0}$ and $F_{1}$ can be expressed as:
\begin{figure}[H]
    \centering 
    \includegraphics[width=0.6\textwidth]{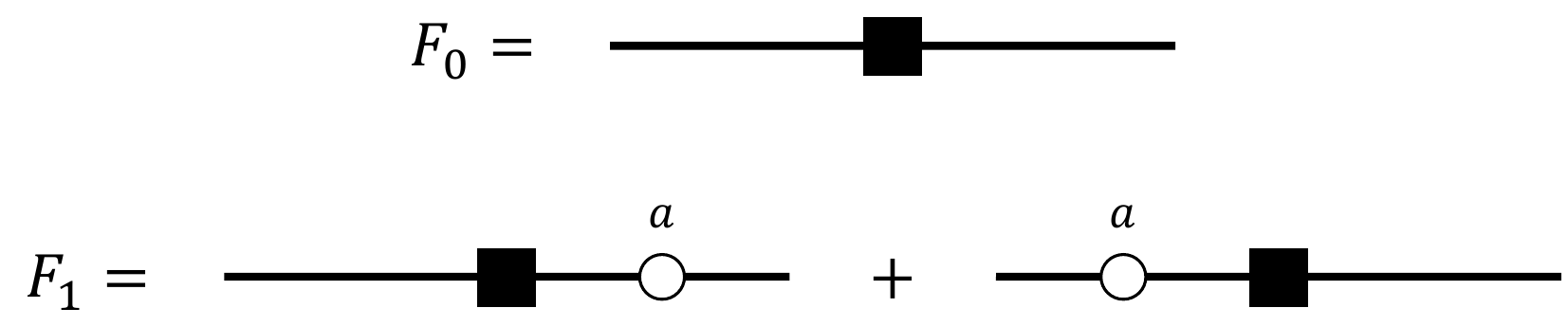}
    \label{fig:diag2}
\end{figure}
As illustrated in the diagrams above, the first-order diagrams confirm that all possible processes, where the weak lens is situated either in front of or behind the strong lens, are summed. Consequently, higher-order amplification factors can be derived by comprehensively summing all conceivable diagrams in a similar manner. 
By considering all possible diagrams consisting of two vertices, we obtain $F_{2}$ as follows

\begin{figure}[H]
    \centering 
    \includegraphics[width=0.8\textwidth]{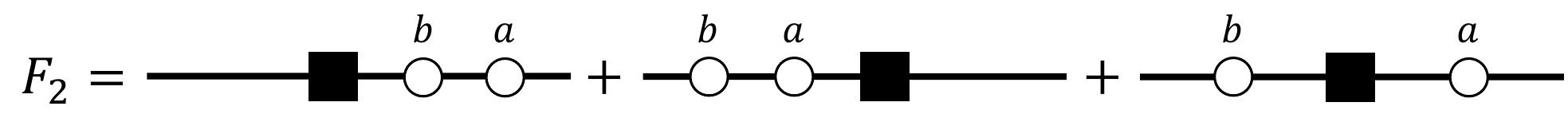}
    \label{fig:F2diag}
\end{figure}
Based on the correspondence between the diagrams and mathematical expressions provided in Figure \ref{fig:diag1} and the rules established in this subsection, the mathematical expression derived from this diagrammatic representation of $F_{2}$ is given by

\begin{align}
    F_{2}&=\int^{\chi_{l}}_{\epsilon}d\chi_{b}\int^{\chi_{b}}_{\epsilon}d\chi_{a}\int d\bm{\theta}_{l}d\bm{\theta}_{b}d\bm{\theta}_{a}d\bm{\theta}_{\epsilon}
    \bra{\bm{\theta}_{s}}\hat{U}_{0}(\chi_{s},\chi_{l})\ket{\bm{\theta}_{l}}\exp{-i\omega\psi(\bm{\theta}_{l})}\bra{\bm{\theta}_{l}}\hat{U}_{0}(\chi_{l},\chi_{b})\ket{\bm{\theta}_{b}}\nonumber\\
    &\times\frac{2\omega}{i}\Phi_{W}(\chi_{b},\bm{\theta}_{b})\bra{\bm{\theta}_{b}}\hat{U}_{0}(\chi_{b},\chi_{a})\ket{\bm{\theta}_{a}}\frac{2\omega}{i}\Phi_{W}(\chi_{a},\bm{\theta}_{a})\bra{\bm{\theta}_{a}}\hat{U}_{0}(\chi_{a},\epsilon)\ket{\bm{\theta}_{\epsilon}}\nonumber\\
    &+\int^{\chi_{s}}_{\chi_{l}}d\chi_{b}\int^{\chi_{b}}_{\chi_{l}}d\chi_{a}\int d\bm{\theta}_{l}d\bm{\theta}_{b}d\bm{\theta}_{a}d\bm{\theta}_{\epsilon}
    \bra{\bm{\theta}_{s}}\hat{U}_{0}(\chi_{s},\chi_{b})\ket{\bm{\theta}_{b}}\frac{2\omega}{i}\Phi_{W}(\chi_{b},\bm{\theta}_{b})\bra{\bm{\theta}_{b}}\hat{U}_{0}(\chi_{b},\chi_{a})\ket{\bm{\theta}_{a}}\nonumber\\
    &\times\frac{2\omega}{i}\Phi_{W}(\chi_{a},\bm{\theta}_{a})\bra{\bm{\theta}_{a}}\hat{U}_{0}(\chi_{a},\chi_{l})\ket{\bm{\theta}_{l}}\exp{-i\omega\psi(\bm{\theta}_{l})}\bra{\bm{\theta}_{l}}\hat{U}_{0}(\chi_{l},\epsilon)\ket{\bm{\theta}_{\epsilon}}\nonumber\\
    &+\int^{\chi_{s}}_{\chi_{l}}d\chi_{b}\int^{\chi_{l}}_{\epsilon}d\chi_{a}\int d\bm{\theta}_{l}d\bm{\theta}_{b}d\bm{\theta}_{a}d\bm{\theta}_{\epsilon}
    \bra{\bm{\theta}_{s}}\hat{U}_{0}(\chi_{s},\chi_{b})\ket{\bm{\theta}_{b}}\frac{2\omega}{i}\Phi_{W}(\chi_{b},\bm{\theta}_{b})\bra{\bm{\theta}_{b}}\hat{U}_{0}(\chi_{b},\chi_{l})\ket{\bm{\theta}_{l}}\nonumber\\
    &\times\exp{-i\omega\psi(\bm{\theta}_{l})}\bra{\bm{\theta}_{l}}\hat{U}_{0}(\chi_{l},\chi_{a})\ket{\bm{\theta}_{a}}\frac{2\omega}{i}\Phi_{W}(\chi_{a},\bm{\theta}_{a})\bra{\bm{\theta}_{a}}\hat{U}_{0}(\chi_{a},\epsilon)\ket{\bm{\theta}_{\epsilon}}.\label{F2}
\end{align}
Eqs. \eqref{F0}, \eqref{F1}, and \eqref{F2} are consistent with the expression derived directly from the path integral formulation in Appendix \ref{apeA}. For the analysis in the following section, we employ the spatial Fourier transform of the gravitational potential:
\begin{align}
    \Phi_{W}(\chi_{a},\bm{\theta}_{a})=\int \frac{d\bm{k}_{a}}{(2\pi)^{3}}\Tilde{\Phi}_{W}(\chi_{a},\bm{k}_{a})e^{ik_{a\parallel}\chi_{a}+i\bm{k}_{a\perp}\cdot \chi_{a}\bm{\theta}_{a}}.\label{Fgrapo}
\end{align}
Here, we reconsider the time dependence of the gravitational potential and replace $\Phi_W(\chi_a,\bm{\theta}_a)$ with $\Phi_W(\eta_a,\chi_a,\bm{\theta}_a)$. Furthermore, the conformal time $\eta$ in the gravitational potential $\Phi_W$ is replaced with the radial distance $\chi$ by using the relation $\eta=\chi$.
Finally, by expressing the gravitational potential $\Phi_{W}$ in terms of its Fourier components $\Tilde{\Phi}_{W}$ using Eq. \eqref{Fgrapo}, the amplification factors up to the second order are summarized below:
\begin{align}
    F_{j0}&=\abs{\mu_{j}}^{1/2}e^{i\omega\Delta t_{S}(\bm{\theta}_{l,j})-i\pi n_{j}\text{sgn}(\omega)}\label{F0f},\\
    \delta F_{j1}\defeq\frac{F_{j1}}{F_{j0}}&=\int^{\chi_{l}}_{0}d\chi_{a}\int\frac{d\bm{k}_{a}}{(2\pi)^{3}}\frac{2\omega}{i}\Tilde{\Phi}_{W}(\chi_{a},\bm{k}_{a})e^{ik_{a\parallel} \chi_{a}+i\bm{k}_{a\perp}\cdot\chi_{a}\bm{\theta}_{l,j}}\nonumber\\ 
    &\times\exp{-i\frac{(\chi_{a} k_{a\perp1})^{2}}{2\omega}\qty(\frac{1}{\tau_{al}}+\frac{\mu_{j,1}}{\tau_{ls}})-i\frac{(\chi_{a} k_{a\perp2})^{2}}{2\omega}\qty(\frac{1}{\tau_{al}}+\frac{\mu_{j,2}}{\tau_{ls}})}\nonumber\\
    &+\int^{\chi_{s}}_{\chi_{l}}d \chi_{a}\int\frac{d\bm{k}_{a}}{(2\pi)^{3}}\frac{2\omega}{i}\Tilde{\Phi}(\chi_{a},\bm{k}_{a})e^{ik_{a\parallel}\chi_{a} +i\bm{k}_{a\perp}\cdot\chi_{a}\qty(\frac{\tau_{la}\bm{\theta}_{l,j}+\tau_{as}\bm{\theta}_{s}}{\tau_{la}+\tau_{as}})}\nonumber\\
    &\times\exp{-i\frac{(\chi_{a}k_{a\perp1})^{2}}{2\omega}\frac{\tau_{ls}}{\tau_{as}}\qty(\frac{1}{\tau_{la}}+\frac{\mu_{j,1}}{\tau_{as}})-i\frac{(\chi_{a}k_{a\perp2})^{2}}{2\omega}\frac{\tau_{ls}}{\tau_{as}}\qty(\frac{1}{\tau_{la}}+\frac{\mu_{j,2}}{\tau_{as}})},\label{F1f}
\end{align}
\begin{align}
    \delta F_{j2}\defeq\frac{F_{j2}}{F_{j0}}&=\int^{\chi_{l}}_{0}d\chi_{b}\int^{\chi_{b}}_{0}d\chi_{a}\int\frac{d\bm{k}_{a}}{(2\pi)^{3}}\int\frac{d\bm{k}_{b}}{(2\pi)^{3}}\frac{2\omega}{i}\Tilde{\Phi}_{W}(\chi_{a},\bm{k}_{a})\frac{2\omega}{i}\Tilde{\Phi}_{W}(\chi_{b},\bm{k}_{b})e^{ik_{a\parallel}\chi_{a}+ik_{b\parallel}\chi_{b}+i(\bm{k}_{a\perp}\chi_{a}+\bm{k}_{b\perp}\chi_{b})\cdot\bm{\theta}_{l,j}}\nonumber\\
    &\times\exp{-i\frac{(\chi_{a}\bm{k}_{a\perp})^{2}}{2\omega\tau_{ab}}-i\frac{(\chi_{a}k_{a\perp1}+\chi_{b}k_{b\perp1})^{2}}{2\omega}\qty(\frac{1}{\tau_{bl}}+\frac{\mu_{j,1}}{\tau_{ls}})-i\frac{(\chi_{a}k_{a\perp2}+\chi_{b}k_{b\perp2})^{2}}{2\omega}\qty(\frac{1}{\tau_{bl}}+\frac{\mu_{j,2}}{\tau_{ls}})}\nonumber\\
    &+\int^{\chi_{s}}_{\chi_{l}}d\chi_{b}\int^{\chi_{b}}_{\chi_{l}}d\chi_{a}\int\frac{d\bm{k}_{a}}{(2\pi)^{3}}\int\frac{d\bm{k}_{b}}{(2\pi)^{3}}\frac{2\omega}{i}\Tilde{\Phi}_{W}(\chi_{a},\bm{k}_{a})\frac{2\omega}{i}\Tilde{\Phi}_{W}(\chi_{b},\bm{k}_{b})e^{ik_{a\parallel}\chi_{a}+ik_{b\parallel}\chi_{b}}\nonumber\\
    &\times\exp{i\chi_{a}\bm{k}_{a\perp}\cdot\frac{\tau_{la}\bm{\theta}_{l,j}+\tau_{as}\bm{\theta}_{s}}{\tau_{la}+\tau_{as}}+i\chi_{b}\bm{k}_{b\perp}\cdot\frac{\tau_{lb}\bm{\theta}_{l,j}+\tau_{bs}\bm{\theta}_{s}}{\tau_{lb}+\tau_{bs}}}\nonumber\\
    &\times\exp\Bigg\{-i\frac{(\chi_{a}\bm{k}_{a\perp})^{2}}{2\omega(\tau_{la}+\tau_{ab})}-\frac{i}{2\omega(\tau_{lb}+\tau_{bs})}\qty(\chi_{b}\bm{k}_{b\perp}+\frac{\tau_{lb}}{\tau_{la}}\chi_{a}\bm{k}_{a\perp})^{2}\nonumber
    \\
    &-i\frac{\mu_{j,1}}{2\omega\tau_{ls}}\qty(\frac{\tau_{ls}}{\tau_{as}}\chi_{a}k_{a\perp1}+\frac{\tau_{ls}}{\tau_{bs}}\chi_{b}k_{b\perp1})^{2}-i\frac{\mu_{j,2}}{2\omega\tau_{ls}}\qty(\frac{\tau_{ls}}{\tau_{as}}\chi_{a}k_{a\perp2}+\frac{\tau_{ls}}{\tau_{bs}}\chi_{b}k_{b\perp2})^{2}\Bigg\}\nonumber\\
    &+\int^{\chi_{s}}_{\chi_{l}}d\chi_{b}\int^{\chi_{l}}_{0}d\chi_{a}\int\frac{d\bm{k}_{a}}{(2\pi)^{3}}\int\frac{d\bm{k}_{b}}{(2\pi)^{3}}\frac{2\omega}{i}\Tilde{\Phi}_{W}(\chi_{a},\bm{k}_{a})\frac{2\omega}{i}\Tilde{\Phi}_{W}(\chi_{b},\bm{k}_{b})e^{ik_{a\parallel}\chi_{a}+ik_{b\parallel}\chi_{b}}
    \nonumber\\
    &\times\exp{i\chi_{a}\bm{k}_{a\perp}\cdot\bm{\theta}_{l,j}+i\chi_{b}\bm{k}_{b\perp}\cdot\frac{\tau_{lb}\bm{\theta}_{l,j}+\tau_{bs}\bm{\theta}_{s}}{\tau_{lb}+\tau_{bs}}}\nonumber\\
    &\times\exp{-i\frac{(\chi_{a}\bm{k}_{a\perp})^{2}}{2\omega\tau_{al}}-i\frac{(\chi_{b}\bm{k}_{b\perp})^{2}}{2\omega(\tau_{lb}+\tau_{bs})}-i\frac{\mu_{j,1}}{2\omega\tau_{ls}}\qty(\chi_{a}k_{a\perp1}+\frac{\tau_{ls}}{\tau_{bs}}\chi_{b}k_{b\perp1})^{2}-i\frac{\mu_{j,2}}{2\omega\tau_{ls}}\qty(\chi_{a}k_{a\perp2}+\frac{\tau_{ls}}{\tau_{bs}}\chi_{b}k_{b\perp2})^{2}},
    \label{F2f}
\end{align}
where $\delta F_{j1}$ and $\delta F_{j2}$ are defined as the ratios of $F_{j1}$ and $F_{j2}$ to $F_{j0}$, respectively. The quantity $\tau_{ab}$ is defined by the following expression:
\begin{align}
    \tau_{ab}\defeq\frac{\chi_{a}\chi_{b}}{\chi_{b}-\chi_{a}}.
\end{align}
Additionally, since the integration over $\bm{\theta}_{\epsilon}$ eliminates the singularity, we have set $\epsilon$ back to zero.

\section{Consistency Relations}
In this section, we demonstrate that the consistency relations reported in previous studies \cite{inamori,mizuno:new} remain valid even in the presence of both strong and cosmological weak lensing, and investigate their physical origin. In Subsection 3.1, we show that the statistical quantities of the fluctuations derived from our expression for the amplification factor, Eqs. \eqref{F0f}, \eqref{F1f} and \eqref{F2f}, also satisfy consistency relations of the same form. Subsequently, Subsections 3.2 and 3.3 are devoted to discussing the physical meaning of these relations. Finally, in Subsection 3.4, we perform numerical calculations to demonstrate that the magnification effect by the strong lens, as identified in prior research \cite{oguri:strong}, occurs within our framework as well.
\subsection{Lensing signal and Consistency relations}

From Eqs. \eqref{F0f}, \eqref{F1f} and \eqref{F2f}, the total amplification factor is summarized as:
\begin{align}
    F_{j}(\omega,\chi_{s})=F_{j0}\qty(1+\delta F_{j1}+\delta F_{j2})+\mathcal{O}(\Phi_{W}^{3}), \label{Fjall}
\end{align}
To characterize the amplitude and phase fluctuations, we generalize the approach in \cite{mizuno:new} and parameterize the amplification factor in the following form:
\begin{align}
    F_{j}(\omega,\chi_{s})\defeq F_{j0}e^{K_{j}+iS_{j}^{\text{tot}}}.\label{Fjdef}
\end{align}
Here, $K_{j}$ and $S_{j}^{\text{tot}}$, both of which are real numbers, denote the amplitude and phase fluctuations, respectively, where $S_{j}^{\text{tot}}$ incorporates the Shapiro time delay—the time lag induced by the gravitational potential. Setting $K_{j}=K_{j1}+K_{j2}$ and $S_{j}^{\text{tot}}=S_{j1}^{\text{tot}}+S_{j2}^{\text{tot}}$, where $K_{j1}$, $S_{j1}^{\text{tot}}$ and $K_{j2}$ , $S_{j2}^{\text{tot}}$ are the first- and second-order terms in $\Phi_{W}$, respectively, a Taylor expansion of the part involving $K_{j}$ and $S_{j}^{\text{tot}}$ on the right-hand side of Eq. \eqref{Fjdef} yields:
\begin{align}
    e^{K_{j}+iS_{j}^{\text{tot}}}=1+K_{j1}+K_{j2}+\frac{K_{j1}^{2}-(S_{j1}^{\text{tot}})^{2}}{2}+i(S_{j1}^{\text{tot}}+S_{j2}^{\text{tot}}+K_{j1}S_{j1}^{\text{tot}})+\mathcal{O}(\Phi_{W}^{3}).\label{expks}
\end{align}
From Eqs. \eqref{Fjall}, \eqref{Fjdef}, and \eqref{expks}, we obtain the following relations:
\begin{align}
    &K_{j1}=\Re\qty(\delta F_{j1}), \qquad S_{j1}^{\text{tot}}=\Im \qty(\delta F_{j1}),\label{ks1} \\
    &K_{j2}+\frac{K_{j1}^{2}-(S_{j1}^{\text{tot}})^{2}}{2}=\Re \qty(\delta F_{j2}),\quad S_{j2}^{\text{tot}}+K_{j1}S_{j1}^{\text{tot}}=\Im \qty(\delta F_{j2}).\label{ks2}
\end{align}
As mentioned earlier, $S^{\text{tot}}_{j}$ includes the Shapiro time delay, which must be removed as it does not constitute an observable quantity. Accordingly, we define the physical phase fluctuation as follows:
\begin{align}
    \frac{S_{j}(\omega)}{\omega}\defeq\frac{S_{j}^{\text{tot}}(\omega)}{\omega}-\lim_{\omega\rightarrow\infty}\frac{S_{j}^{\text{tot}}(\omega)}{\omega}.\label{shapi}
\end{align}
By using Eqs. \eqref{F0f}-\eqref{F2f} and \eqref{ks1}-\eqref{shapi}, we obtain the explicit expressions for $K_{j1}$, $K_{j2}$, $S_{j1}$, and $S_{j2}$, which are summarized in Appendix \ref{apeB}. The explicit expressions for the first- and second-order Shapiro time delays subtracted in this calculation are also summarized in Appendix \ref{apeB}.

These expressions provide the deterministic form of the amplification factor for a specific realization of the gravitational potential. However, to relate these theoretical results to actual cosmological observations, we must account for the stochastic nature of the universe. 
The cosmological matter inhomogeneity causing the weak lensing potential $\Phi_{W}$ is described as a statistical random field. Therefore, to compare our derived result with observational data, it is necessary to evaluate its statistical average. Since the statistical mean of the gravitational potential vanishes, under the global homogeneity and isotropy of the universe, we use the power spectrum, defined as follows:
\begin{align}
    \big\langle\Tilde{\Phi}_{W}(\chi_{a},\bm{k}_{a})\Tilde{\Phi}_{W}(\chi_{b},\bm{k}_{b})\big\rangle_{W}=(2\pi)^{3}\delta^{3}(\bm{k}_{a}+\bm{k}_{b})P_{\Phi}(\chi_{a},\chi_{b},k_{a}). \label{gravpo}
\end{align}
where $\langle \cdots\rangle_{W}$ and $P_{\Phi}$ denote the statistical average with respect to the weak lensing potential and the power spectrum of the gravitational potential, respectively. Since we are considering a single fixed strong lens, this statistical average acts only on the weak lensing potential and does not affect the strong lensing potential. 

Following the procedure in \cite{oguri:probe}, we now calculate the mean and variance of $K_{j}$ and $S_{j}$ up to the second order in $\Phi_{W}$, using the average of $\Phi_{W}$ defined in Eq. \eqref{gravpo} and the Limber approximation. Using Eqs. \eqref{kj1spe} and \eqref{sj1spe}, the variances of $K_{j}$ and $S_{j}$ are given by
\begin{align}
    \langle K^{2}_{j}\rangle_{W}=&4\omega^{2}\int^{\chi_{l}}_{0}d\chi_{a}\int\frac{d\bm{k}_{a\perp}}{(2\pi)^{2}}P_{\Phi}(\chi_{a},k_{a\perp})\sin^{2}{\qty{\qty(\frac{1}{\tau_{al}}+\frac{\mu_{j,1}}{\tau_{ls}})\frac{(\chi_{a}k_{a\perp1})^{2}}{2\omega}+\qty(\frac{1}{\tau_{al}}+\frac{\mu_{j,2}}{\tau_{ls}})\frac{(\chi_{a}k_{a\perp2})^{2}}{2\omega}}}\nonumber\\
    +&4\omega^{2}\int^{\chi_{s}}_{\chi_{l}}d\chi_{a}\int\frac{d\bm{k}_{a\perp}}{(2\pi)^{2}}P_{\Phi}(\chi_{a},k_{a\perp})\sin^{2}{\qty{\frac{\tau_{ls}}{\tau_{as}}\qty(\frac{1}{\tau_{la}}+\frac{\mu_{j,1}}{\tau_{as}})\frac{(\chi_{a}k_{a\perp1})^{2}}{2\omega}+\frac{\tau_{ls}}{\tau_{as}}\qty(\frac{1}{\tau_{la}}+\frac{\mu_{j,2}}{\tau_{as}})\frac{(\chi_{a}k_{a\perp2})^{2}}{2\omega}}},\label{<k2>}\\
    \langle S^{2}_{j}\rangle_{W}=&4\omega^{2}\int^{\chi_{l}}_{0}d\chi_{a}\int\frac{d\bm{k}_{a\perp}}{(2\pi)^{2}}P_{\Phi}(\chi_{a},k_{a\perp})\Bigg[1-\cos{\qty{\qty(\frac{1}{\tau_{al}}+\frac{\mu_{j,1}}{\tau_{ls}})\frac{(\chi_{a}k_{a\perp1})^{2}}{2\omega}+\qty(\frac{1}{\tau_{al}}+\frac{\mu_{j,2}}{\tau_{ls}})\frac{(\chi_{a}k_{a\perp2})^{2}}{2\omega}}}\Bigg]^{2}\nonumber\\
    +&4\omega^2\int^{\chi_{s}}_{\chi_{l}}d\chi_{a}\int\frac{d\bm{k}_{a\perp}}{(2\pi)^{2}}P_{\Phi}(\chi_{a},k_{a\perp})\Bigg[1-\cos{\qty{\frac{\tau_{ls}}{\tau_{as}}\qty(\frac{1}{\tau_{la}}+\frac{\mu_{j,1}}{\tau_{as}})\frac{(\chi_{a}k_{a\perp1})^{2}}{2\omega}+\frac{\tau_{ls}}{\tau_{as}}\qty(\frac{1}{\tau_{la}}+\frac{\mu_{j,2}}{\tau_{as}})\frac{(\chi_{a}k_{a\perp2})^{2}}{2\omega}}}\Bigg]^{2}.\label{<s2>}
\end{align}
Here, we have used the fact that the terms involving $\int^{\chi_{s}}_{\chi_{l}}d\chi_{b}\int^{\chi_{l}}_{0}d\chi_{a}\delta(\chi_{a}-\chi_{b})$ vanish because the integration regions for $\chi_{a}$ and $\chi_{b}$ do not overlap, except at the boundary point $\chi_{l}$.
Furthermore, using Eq. \eqref{kj1spe} and \eqref{sj1spe}, the correlation $\langle K_{j}S_{j}\rangle$ is also obtained as
\begin{align}
    \langle K_{j}S_{j}\rangle_{W}&=2\omega^{2}\int^{\chi_{l}}_{0}d\chi_{a}\int\frac{d\bm{k}_{a\perp}}{(2\pi)^{2}}P_{\Phi}(\chi_{a},k_{a\perp})\Bigg[\sin{\qty{\qty(\frac{1}{\tau_{al}}+\frac{\mu_{j,1}}{\tau_{ls}})\frac{(\chi_{a}k_{a\perp1})^{2}}{\omega}+\qty(\frac{1}{\tau_{al}}+\frac{\mu_{j,2}}{\tau_{ls}})\frac{(\chi_{a}k_{a\perp2})^{2}}{\omega}}}\nonumber\\
    &-2\sin{\qty{\qty(\frac{1}{\tau_{al}}+\frac{\mu_{j,1}}{\tau_{ls}})\frac{(\chi_{a}k_{a\perp1})^{2}}{2\omega}+\qty(\frac{1}{\tau_{al}}+\frac{\mu_{j,2}}{\tau_{ls}})\frac{(\chi_{a}k_{a\perp2})^{2}}{2\omega}}}\Bigg]\nonumber\\
    &+2\omega^{2}\int^{\chi_{s}}_{\chi_{l}}d\chi_{a}\int\frac{d\bm{k}_{a\perp}}{(2\pi)^{2}}P_{\Phi}(\chi_{a},k_{a\perp})\Bigg[\sin{\qty{\frac{\tau_{ls}}{\tau_{as}}\qty(\frac{1}{\tau_{la}}+\frac{\mu_{j,1}}{\tau_{as}})\frac{(\chi_{a}k_{a\perp1})^{2}}{\omega}+\frac{\tau_{ls}}{\tau_{as}}\qty(\frac{1}{\tau_{la}}+\frac{\mu_{j,2}}{\tau_{as}})\frac{(\chi_{a}k_{a\perp2})^{2}}{\omega}}}\nonumber\\
    &-2\sin{\qty{\frac{\tau_{ls}}{\tau_{as}}\qty(\frac{1}{\tau_{la}}+\frac{\mu_{j,1}}{\tau_{as}})\frac{(\chi_{a}k_{a\perp1})^{2}}{2\omega}+\frac{\tau_{ls}}{\tau_{as}}\qty(\frac{1}{\tau_{la}}+\frac{\mu_{j,2}}{\tau_{as}})\frac{(\chi_{a}k_{a\perp2})^{2}}{2\omega}}}\Bigg].\label{<ks>}
\end{align}
While the quantities above are derived from the first-order term in $\Phi_{W}$ of the amplification factor, the means of $K_{j}$ and $S_{j}$ are obtained from the second-order term, as the first-order means such as $\langle K_{j1}\rangle$ vanish due to $\langle\Phi_{W}\rangle=0$. By using Eqs. \eqref{kj2spe} and \eqref{sj2spe}, along with the statistical quantities already derived in Eqs. \eqref{<k2>}, \eqref{<s2>}, and \eqref{<ks>}, the means of $K_{j}$ and $S_{j}$ are calculated as follows:
\begin{align}
    \langle K_{j}\rangle_{W}=&-4\omega^{2}\int^{\chi_{l}}_{0}d\chi_{a}\int\frac{d\bm{k}_{a\perp}}{(2\pi)^{2}}P_{\Phi}(\chi_{a},k_{a\perp})\sin^{2}{\qty{\qty(\frac{1}{\tau_{al}}+\frac{\mu_{j,1}}{\tau_{ls}})\frac{(\chi_{a}k_{a\perp1})^{2}}{2\omega}+\qty(\frac{1}{\tau_{al}}+\frac{\mu_{j,2}}{\tau_{ls}})\frac{(\chi_{a}k_{a\perp2})^{2}}{2\omega}}}\nonumber\\
    &-4\omega^{2}\int^{\chi_{s}}_{\chi_{l}}d\chi_{a}\int\frac{d\bm{k}_{a\perp}}{(2\pi)^{2}}P_{\Phi}(\chi_{a},k_{a\perp})\sin^{2}{\qty{\frac{\tau_{ls}}{\tau_{as}}\qty(\frac{1}{\tau_{la}}+\frac{\mu_{j,1}}{\tau_{as}})\frac{(\chi_{a}k_{a\perp1})^{2}}{2\omega}+\frac{\tau_{ls}}{\tau_{as}}\qty(\frac{1}{\tau_{la}}+\frac{\mu_{j,2}}{\tau_{as}})\frac{(\chi_{a}k_{a\perp2})^{2}}{2\omega}}},\label{<k>}
    \\
    \langle S_{j}\rangle_{W}&=2\omega^{2}\int^{\chi_{l}}_{0}d\chi_{a}\int\frac{d\bm{k}_{a\perp}}{(2\pi)^{2}}P_{\Phi}(\chi_{a},k_{a\perp})\Bigg[\qty{\qty(\frac{1}{\tau_{al}}+\frac{\mu_{j,1}}{\tau_{ls}})\frac{(\chi_{a}k_{a\perp1})^{2}}{\omega}+\qty(\frac{1}{\tau_{al}}+\frac{\mu_{j,2}}{\tau_{ls}})\frac{(\chi_{a}k_{a\perp2})^{2}}{\omega}}\nonumber\\
    &-\sin{\qty{\qty(\frac{1}{\tau_{al}}+\frac{\mu_{j,1}}{\tau_{ls}})\frac{(\chi_{a}k_{a\perp1})^{2}}{\omega}+\qty(\frac{1}{\tau_{al}}+\frac{\mu_{j,2}}{\tau_{ls}})\frac{(\chi_{a}k_{a\perp2})^{2}}{\omega}}}\Bigg]\nonumber\\
    &+2\omega^{2}\int^{\chi_{s}}_{\chi_{l}}d\chi_{a}\int\frac{d\bm{k}_{a\perp}}{(2\pi)^{2}}P_{\Phi}(\chi_{a},k_{a\perp})\Bigg[\qty{\frac{\tau_{ls}}{\tau_{as}}\qty(\frac{1}{\tau_{la}}+\frac{\mu_{j,1}}{\tau_{as}})\frac{(\chi_{a}k_{a\perp1})^{2}}{\omega}+\frac{\tau_{ls}}{\tau_{as}}\qty(\frac{1}{\tau_{la}}+\frac{\mu_{j,2}}{\tau_{as}})\frac{(\chi_{a}k_{a\perp2})^{2}}{\omega}}\nonumber\\
    &-\sin{\qty{\frac{\tau_{ls}}{\tau_{as}}\qty(\frac{1}{\tau_{la}}+\frac{\mu_{j,1}}{\tau_{as}})\frac{(\chi_{a}k_{a\perp1})^{2}}{\omega}+\frac{\tau_{ls}}{\tau_{as}}\qty(\frac{1}{\tau_{la}}+\frac{\mu_{j,2}}{\tau_{as}})\frac{(\chi_{a}k_{a\perp2})^{2}}{\omega}}}\Bigg].\label{<s>}
\end{align}
Details regarding the handling of delta functions in the above derivation are provided in Appendix \ref{apeB}.
Using Eqs. \eqref{<k2>}, \eqref{<s2>}, \eqref{<ks>}, \eqref{<k>}, and \eqref{<s>}, we can check that our formulae satisfy the following relations:
\begin{align}
    &\langle K^{2}_{j}(\omega)\rangle_{W}+\langle S^{2}_{j}(\omega)\rangle_{W}=\langle K^{2}_{j}(2\omega)\rangle_{W},\label{cons1}\\
    &\langle S_{j}(\omega)\rangle_{W}-\frac{1}{2}\langle S_{j}(2\omega)\rangle_{W}=-\langle K_{j}(\omega)S_{j}(\omega)\rangle_{W},\label{cons2}\\
    &\langle K^{2}_{j}(\omega)\rangle_{W}+\langle K_{j}(\omega)\rangle_{W}=0.\label{cons3}
\end{align}
These are precisely the consistency relations, which represent the most significant result of this work. By demonstrating that these relations hold, we have confirmed that they can serve as a robust observational test even in the general scenario involving both strong and weak lenses, as considered in this study. It should be noted, however, that the physical context of the statistical average differs from previous works: while the average in \cite{inamori, mizuno:new} operates on the general gravitational potential, our average acts exclusively on the weak lensing potential and does not affect the strong lens, as stated above.

Furthermore, by using Eq. \eqref{cons3}, Eqs. \eqref{cons1} and \eqref{cons2} can be combined into the following form:
\begin{align}
\langle K_{j}(\omega)+iS_{j}(\omega)\rangle_{W}-\frac{1}{2}\langle K_{j}(2\omega)+iS_{j}(2\omega)\rangle_{W}=-\frac{1}{2}\Big\langle(K_{j}(\omega)+iS_{j}(\omega))^{2}\Big\rangle_{W}.\label{cons12}
\end{align}

\subsection{Shapiro time delay and Consistency relations}
Next, we discuss the physical interpretation of the consistency relations. In previous literature \cite{mizuno:new}, it has been pointed out that some of the consistency relations originate from the Shapiro time delay. In this subsection, we demonstrate that a similar interpretation holds within our framework.
We begin with the path-integral representation of the amplification factor \cite{Nakamura}:
\begin{align}
    F(\omega,\chi_{s},\bm{\theta}_{s})&=\int \mathscr{D}\bm{\theta}\exp{i\omega\int^{\chi_{s}}_{0} d\chi\left[ \frac{1}{2}\chi^{2}|\dot{\bm{\theta}}|^{2}-\psi(\bm{\theta}_l)\delta(\chi-\chi_l)-2\Phi_{W}(\chi,\bm{\theta})\right]}.\label{Fpath}
\end{align}
where the integration measure is normalized such that $F=1$ in the absence of any gravitational potential $\Phi=0$. Taking the ensemble average $\langle\cdots\rangle_{W}$ with respect to the weak lensing potential yields:
\begin{align}
    \langle F(\omega,\chi_{s},\bm{\theta}_{s}) \rangle_{W}=\int \mathscr{D}\bm{\theta}\exp{i\omega\int^{\chi_{s}}_{0} d\chi\left[ {\frac{1}{2}\chi^{2}|\dot{\bm{\theta}}|^{2}-\psi(\bm{\theta}_l)\delta(\chi-\chi_l)}\right]}\times\Bigg\langle \exp{-2i\omega\int^{\chi_{s}}_{0}d\chi \Phi_{W}(\chi,\bm{\theta})} \Bigg\rangle_{W}.
\end{align}
By applying the Limber approximation, we obtain $\big\langle \exp{-2i\omega\int^{\chi_{s}}_{0}d\chi \Phi_{W}(\chi,\bm{\theta})} \big\rangle_{W}=\big\langle \exp{-2i\omega\int^{\chi_{s}}_{0}d\chi \Phi_{W}(\chi,\bm{\theta}_{s})} \big\rangle_{W}$ \cite{mizuno:new}. As a result, the path integral depends only on the properties of the strong lens, becoming identical in form to Eq. \eqref{F0}. Focusing on the $j$-th image, we obtain:
\begin{align}
    \langle F_{j}(\omega,\chi_{s},\bm{\theta}_{s})\rangle_{W}=\abs{\mu_{j}}^{1/2}e^{i\omega\Delta t_{S}(\bm{\theta}_{l,j})-i\pi n_{j}\text{sgn}(\omega)}\times\Bigg\langle \exp{-2i\omega\int^{\chi_{s}}_{0}d\chi \Phi_{W}(\chi,\bm{\theta}_{s})} \Bigg\rangle_{W}. \label{<Fj>0}
\end{align}
By reapplying the Limber approximation and using the expression for the first-order Shapiro time delay in Eq. \eqref{shapi1} along with Eq. \eqref{F0f}, Eq. \eqref{<Fj>0} can be eventually transformed as follows:
\begin{align}
    \langle F_{j}(\omega,\chi_{s},\bm{\theta}_{s})\rangle_{W}=F_{j0}\times\Big\langle e^{i\omega\Delta t^{(1)}_{Wj}} \Big\rangle_{W}. \label{<Fj>}
\end{align}
Taking the statistical average of both sides of Eq. \eqref{Fjdef}, we obtain the following:
\begin{align}
    \langle F_{j}(\omega,\chi_{s},\bm{\theta}_{s})\rangle_{W}=F_{j0}\times\Big\langle e^{K_{j}}e^{i(S_{j}+\omega\Delta t_{Wj})}\Big\rangle_{W},\label{<Fj>2}
\end{align}
where $\Delta t_{Wj}$ is the Shapiro time delay defined as $\Delta t_{Wj}\defeq\lim_{\omega\rightarrow\infty}S^{tot}_{j}(\omega)/\omega$. From Eqs. \eqref{<Fj>} and \eqref{<Fj>2}, we obtain the following relation:
\begin{align}
    \Big\langle e^{K_{j}}e^{i(S_{j}+\omega\Delta t_{Wj})}\Big\rangle_{W}=\Big\langle e^{i\omega\Delta t^{(1)}_{Wj}} \Big\rangle_{W}.\label{per.con.re}
\end{align}
This result represents the full-order consistency relations, which shares the same functional form as those presented in previous study \cite{mizuno:new}. However, it should be noted that the definition of the statistical average differs; in our context, the average is taken specifically over the weak lensing potential.

To verify whether the consistency relations derived in the previous subsection can be recovered from this full-order result, we first expand Eq. \eqref{per.con.re} up to the second order in $\Phi$. The expanded form is given as follows:
\begin{align}
    &1+\langle K_{j}(\omega)+i(S_{j}(\omega)+\omega\Delta t_{Wj})\rangle_{W}+\frac{1}{2}\Big\langle(K_{j}(\omega)+i(S_{j}(\omega)+\omega\Delta t_{Wj}))^{2}\Big\rangle_{W}+\mathcal{O}(\Phi^{3})\nonumber\\
    =&1+\frac{1}{2}\Big\langle\qty(i\omega\Delta t^{(1)}_{Wj})^{2}\Big\rangle+\mathcal{O}(\Phi^{3}),\label{kst}
\end{align}
where we have used $\langle \Delta t^{(1)}_{Wj}\rangle=0$ since $\langle\Phi\rangle=0$.
Following the procedure described in the previous subsection and Appendix \ref{apeB}, we decompose the amplitude and phase fluctuations as $K_{j}=K_{j1}+K_{j2}$ and $S_{j}=S_{j1}+S_{j2}$, and the Shapiro time delay as $\Delta t_{Wj}=\Delta t^{(1)}_{Wj}+\Delta t^{(2)}_{Wj}$. By substituting the explicit expressions for these quantities given in Eqs. \eqref{kj1spe}-\eqref{shapi2} into Eq. \eqref{kst}, we eventually obtain the following:
\begin{align}
    \langle K_{j}(\omega)+iS_{j}(\omega)\rangle_{W}-\frac{1}{2}\langle K_{j}(2\omega)+iS_{j}(2\omega)\rangle_{W}=-\frac{1}{2}\Big\langle(K_{j}(\omega)+iS_{j}(\omega))^{2}\Big\rangle_{W}.\label{cons12r}
\end{align}
This is precisely Eq. \eqref{cons12}, and hence also Eqs. \eqref{cons1} and \eqref{cons2}.
Therefore, we can conclude that the following physical interpretation remains valid even in the general scenario considered in this study. Specifically, as indicated by Eq. \eqref{Fpath}, the amplification factor is expressed as a superposition of all possible paths from the source to the observer. Gravitational waves along different paths experience varying phase shifts due to the presence of the local weak lensing potential $\Phi_{W}$. When these phase-shifted waves are summed, their interference results in modifications not only to the phase but also to the amplitude. This indicates a non-trivial relationship between phase and amplitude fluctuations, originating from the Shapiro time delay. This relationship manifests as the consistency relations \eqref{cons12r} upon taking the statistical average.

\subsection{Energy conservation law and Consistency relations}
Next, we show that the origin of the final consistency relation \eqref{cons3} lies in the energy conservation law. Since previous studies demonstrated this origin by establishing the relation $\langle \abs{F}^{2}\rangle=1$, we also proceed to evaluate $\langle \abs{F}^{2}\rangle_{W}$. Similarly to the discussion in Subsection 3.2, by employing the path-integral representation of the amplification factor, $\langle \abs{F}^{2}\rangle_{W}$ can be expressed as follows:
\begin{align}
    \langle\abs{F}^{2}\rangle_{W}&=\int\mathscr{D}\bm{\theta}_{1}(\mathscr{D}\bm{\theta}_{2})^{*}\exp{i\omega\int^{\chi_{s}}_{0}d\chi\frac{1}{2}\chi^{2}(|\dot{\bm{\theta}}_{1}|^{2}-|\dot{\bm{\theta}}_{2}|^{2})-i\omega\qty(\psi\qty(\bm{\theta}_{1,l})-\psi\qty(\bm{\theta}_{2,l}))}\nonumber\\
    &\times\Bigg\langle\exp{-2i\omega\int^{\chi_{s}}_{0}d\chi(\Phi_{W}(\chi,\bm{\theta}_{1})-\Phi_{W}(\chi,\bm{\theta}_{2}))}\Bigg\rangle_{W}.\label{F2path}
\end{align}
By applying the stationary phase approximation and the Limber approximation, Eq. \eqref{F2path} can be rewritten as (see Appendix \ref{apeC})
\begin{align}
    \langle\abs{F_{j}}^{2}\rangle_{W}=\abs{\mu_{j}}.\label{Fmu3.26}
\end{align}
In deriving this equation, we reiterate that only the $j$-th image is taken into account.
On the other hand, the expression obtained by substituting Eq. \eqref{Fjdef} into $\langle \abs{F_{j}}^{2}\rangle_{W}$ is given by
\begin{align}
    \langle\abs{F_{j}}^{2}\rangle_{W}=\abs{\mu_{j}}\langle e^{2K_{j}}\rangle_{W}.\label{Fdef3.27}
\end{align}
From Eqs. \eqref{Fmu3.26} and \eqref{Fdef3.27}, we obtain the following relation:
\begin{align}
    \langle e^{2K_{j}}\rangle_{W}=1.
\end{align}
This is precisely the full-order consistency relation originating from the energy conservation law. Expanding this expression up to the second order in $K_{j}$, we obtain:
\begin{align}
    \langle K^{2}_{j}\rangle_{W}+\langle K_{j}\rangle_{W}=0
\end{align}
This is consistent with Eq. \eqref{cons3}, and we can conclude that it originates from the energy conservation law.

\subsection{Dependence of kernel functions on magnifications factors}

Previous study \cite{oguri:strong} has shown that the kernel function, which links the phase fluctuation of gravitational waves to the matter power spectrum, is sensitive to the magnification factor, $\mu_{j}$. In strong gravitational lensing events with a large $\mu_{j}$, the effective Fresnel scale shifts toward larger scales, and the amplitude of the kernel function at that scale is boosted. Consequently, the contribution from the power spectrum around the Fresnel scale is boosted, enhancing the sensitivity to probe matter fluctuations on that scale. In this subsection, we demonstrate numerically that the formulae we have derived also exhibit this property.

In Fourier space, the local convergence $\Tilde{\kappa}^{j}_{\text{loc}}(\chi,\bm{k})$, which represents the density fluctuations, is defined by:
\begin{align}
    -k^{2}\Tilde{\Phi}_{W}(\chi,\bm{k})=\frac{\chi_{s}}{\chi(\chi_{s}-\chi)}\Tilde{\kappa}^{j}_{\text{loc}}(\chi,\bm{k}).\label{phieqkap}
\end{align}
In real space, the standard convergence $\kappa^j(\bm{\theta})$ is obtained by integrating $\kappa^{j}_{\text{loc}}(\chi,\bm{\theta})$ along the line of sight: $\kappa^j(\bm{\theta})=\int^{\chi_s}_{0}d\chi\ \kappa^j_{\text{loc}}(\chi,\bm{\theta})$.
Using the definition of $\Tilde{\kappa}^j_{\text{loc}}$ \eqref{phieqkap}, the relation between the power spectrum of the gravitational potential $P_{\Phi}$ and the local convergence power spectrum $P_{\kappa,\text{loc}}^{j}$ is given by:
\begin{align}
    P_{\Phi}(\chi_{a},\chi_{b},k_a)=\frac{1}{k^{4}_a}\frac{\chi_{s}}{\chi_{a}(\chi_{s}-\chi_{a})}\frac{\chi_{s}}{\chi_{b}(\chi_{s}-\chi_{b})}P^{j}_{\kappa,\text{loc}}(\chi_{a},\chi_{b},k_{a}).\label{pppk}
\end{align}
where the local convergence power spectrum is defined as follows:
\begin{align}
    \langle \Tilde{\kappa}^{j}_{\text{loc}}(\chi_{a},\bm{k}_{a})\Tilde{\kappa}^{j}_{\text{loc}}(\chi_{b},\bm{k}_{b})\rangle_{W}\defeq(2\pi)^{3}\delta^{3}(\bm{k}_{a}+\bm{k}_{b})P^{j}_{\kappa,\text{loc}}(\chi_{a},\chi_{b},k_{a}).\label{pkapa}
\end{align}
Let us define the following quantities:
\begin{align}
    r_{F}(\chi_{a})=\sqrt{\frac{\chi_{a}(\chi_{s}-\chi_{a})}{\omega\chi_{s}}}, \quad &A=\frac{r_{F}^{2}(\chi_{a})k_{a\perp}^{2}}{2}, \quad  A^{j,\pm}(\chi_{a}<\chi_{l})=\qty[\qty(\frac{1}{\tau_{al}}+\frac{\mu_{j,1}}{\tau_{ls}})\pm\qty(\frac{1}{\tau_{al}}+\frac{\mu_{j,2}}{\tau_{ls}})]\frac{(\chi_{a}k_{a\perp})^{2}}{2\omega},\nonumber \\ &A^{j,\pm}(\chi_{a}>\chi_{l})=\qty[\frac{\tau_{ls}}{\tau_{as}}\qty(\frac{1}{\tau_{la}}+\frac{\mu_{j,1}}{\tau_{as}})\pm\frac{\tau_{ls}}{\tau_{as}}\qty(\frac{1}{\tau_{la}}+\frac{\mu_{j,2}}{\tau_{as}})]\frac{(\chi_{a}k_{a\perp})^{2}}{2\omega},
\end{align}
where $r_{F}$ is the Fresnel scale.

\begin{figure}[t!]
    \centering 

    \begin{subfigure}[b]{0.495\textwidth}
        \centering
        \includegraphics[
            trim=0cm 0cm 1.5cm 0cm, 
            clip,
            width=\textwidth
        ]{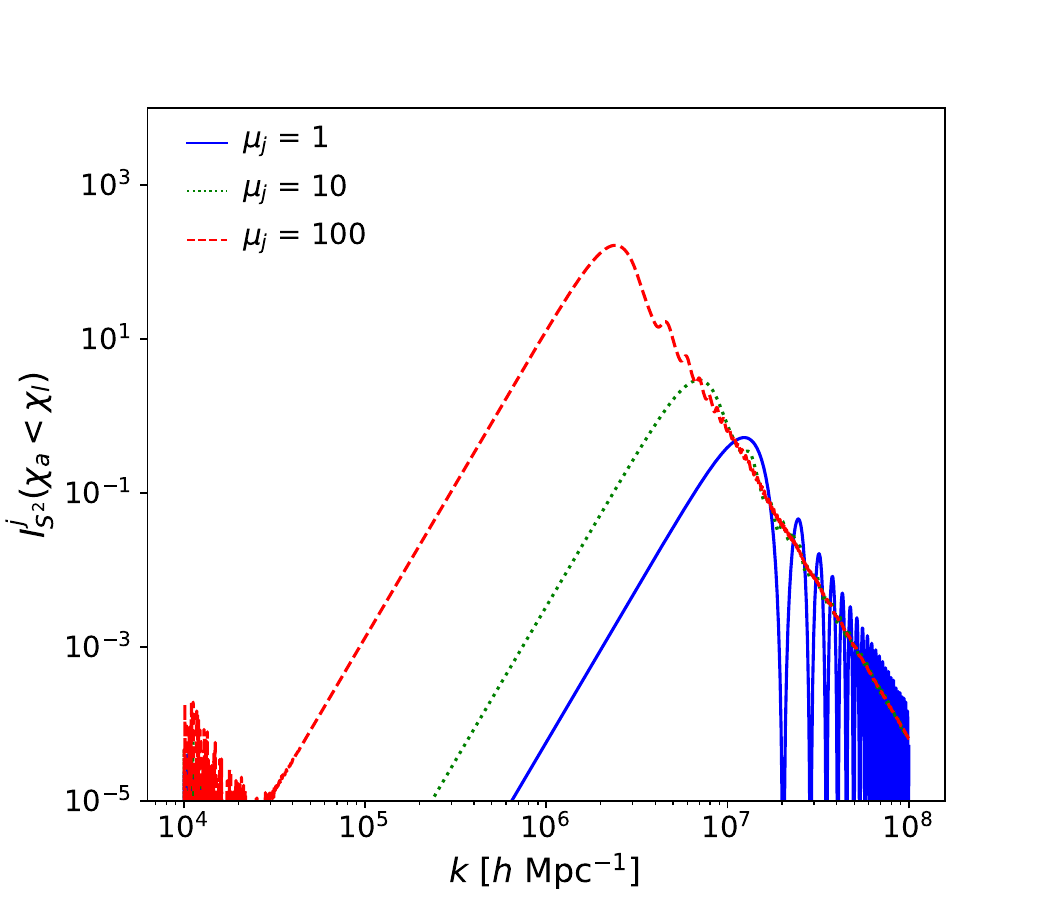}
    \end{subfigure}
    \hfill 
    \begin{subfigure}[b]{0.495\textwidth}
        \centering
        \includegraphics[
            trim=0cm 0cm 1.5cm 0cm, 
            clip,
            width=\textwidth
        ]{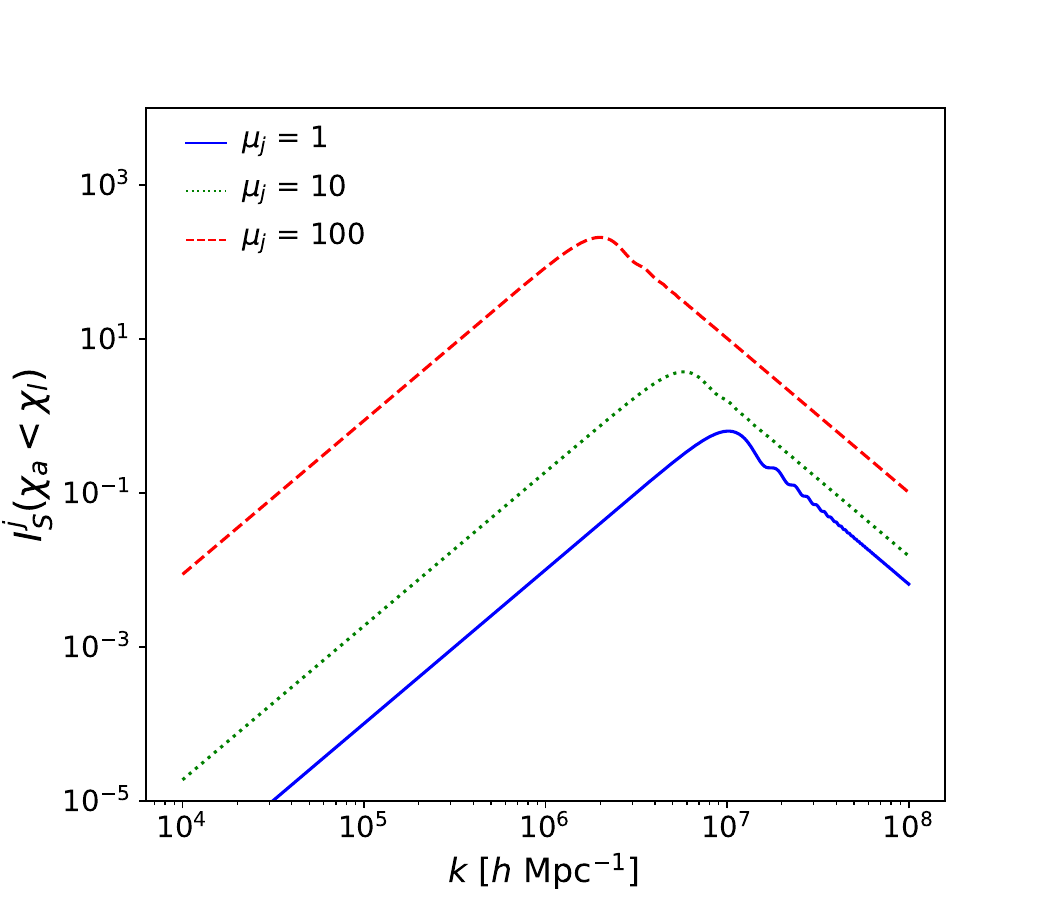}
    \end{subfigure}

    \caption{The kernel functions of $\langle S^{2}_{j}\rangle_{W}$ (left) and $\langle S_{j}\rangle_{W}$ (right) are plotted for three different values of the magnification factor $\mu_{j}=\mu_{j,1}\mu_{j,2}=1\ (\text{solid}), 10\ (\text{dotted}), 100\ (\text{dashed})$. The variation in $\mu_{j}$ is achieved by changing $\mu_{j,1}$ while holding $\mu_{j,2}=1$ fixed. We assume a weak lens redshift of $z=0.1$, a strong lens redshift of $z_{l}=0.3$, a source redshift of $z_{s}=2$, and a gravitational wave frequency of $10\ \text{Hz}$, for which the inverse of the
    Fresnel scale is $1/ r_{F}\sim 6\times10^{6}h \text{Mpc}^{-1}$.}
    \label{fig:main}
\end{figure}

Then, substituting Eq. \eqref{pppk} into Eqs. \eqref{<k2>} and \eqref{<s2>} and performing the angular integration in $k$-space, we obtain $\langle K^{2}_{j}\rangle_{W}$ and $\langle S^{2}_{j}\rangle_{W}$ as follows:
\begin{align}
    \langle K^{2}_{j}\rangle_{W}&=\int^{\chi_{l}}_{0}d\chi_{a}\int\frac{dk_{a\perp}k_{a\perp}}{2\pi}P^{j}_{\kappa,\text{loc}}(\chi_{a},k_{a\perp})I^{j}_{K^{2}}(\chi_{a}<\chi_{l})+\int^{\chi_{s}}_{\chi_{l}}d\chi_{a}\int\frac{dk_{a\perp}k_{a\perp}}{2\pi}P^{j}_{\kappa,\text{loc}}(\chi_{a},k_{a\perp})I^{j}_{K^{2}}(\chi_{a}>\chi_{l}),\\
    \langle S^{2}_{j}\rangle_{W}&=\int^{\chi_{l}}_{0}d\chi_{a}\int\frac{dk_{a\perp}k_{a\perp}}{2\pi}P^{j}_{\kappa,\text{loc}}(\chi_{a},k_{a\perp})I^{j}_{S^{2}}(\chi_{a}<\chi_{l})+\int^{\chi_{s}}_{\chi_{l}}d\chi_{a}\int\frac{dk_{a\perp}k_{a\perp}}{2\pi}P^{j}_{\kappa,\text{loc}}(\chi_{a},k_{a\perp})I^{j}_{S^{2}}(\chi_{a}>\chi_{l}),\\
    \langle K_{j}S_{j}\rangle_{W}&=\int^{\chi_{l}}_{0}d\chi_{a}\int\frac{dk_{a\perp}k_{a\perp}}{2\pi}P^{j}_{\kappa,\text{loc}}(\chi_{a},k_{a\perp})I^{j}_{KS}(\chi_{a}<\chi_{l})+\int^{\chi_{s}}_{\chi_{l}}d\chi_{a}\int\frac{dk_{a\perp}k_{a\perp}}{2\pi}P^{j}_{\kappa,\text{loc}}(\chi_{a},k_{a\perp})I^{j}_{KS}(\chi_{a}>\chi_{l})
\end{align}

\begin{figure}[t!]
    \centering 

    \begin{subfigure}[b]{0.495\textwidth}
        \centering
        \includegraphics[
            trim=0cm 0cm 1.5cm 0cm, 
            clip,
            width=\textwidth
        ]{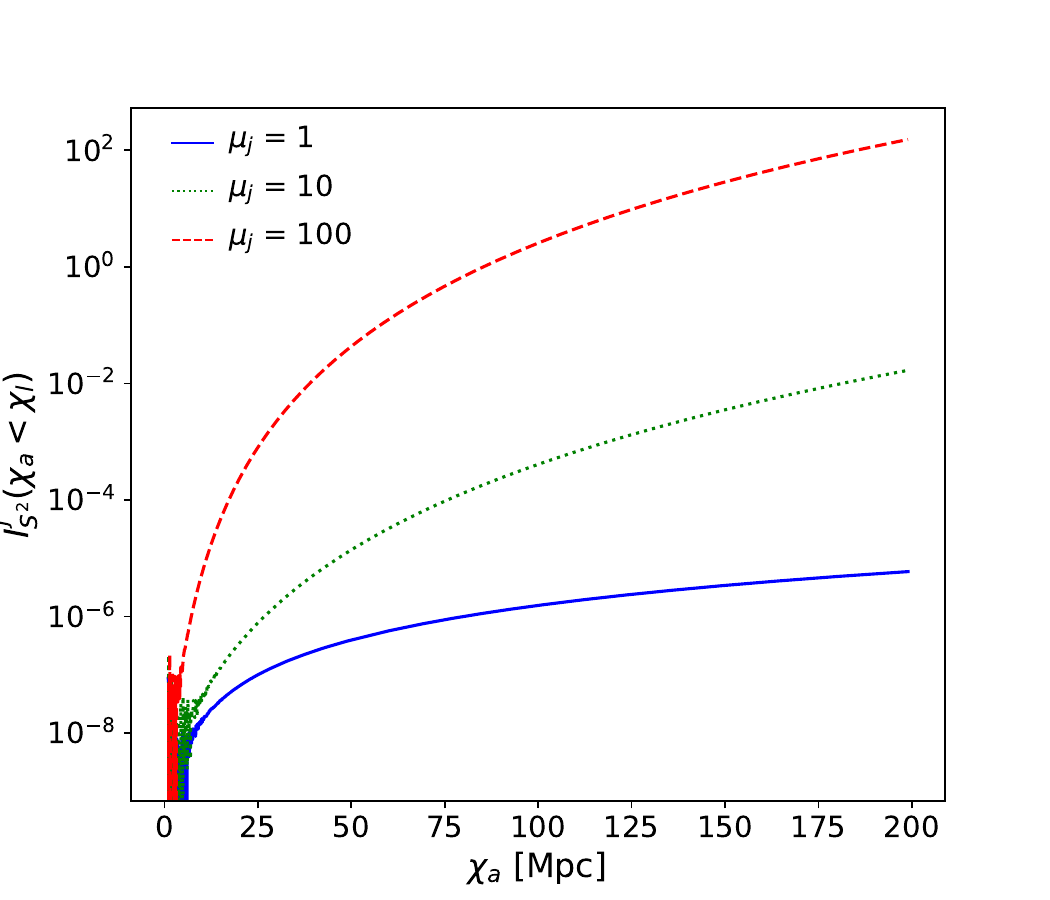}
    \end{subfigure}
    \hfill 
    \begin{subfigure}[b]{0.495\textwidth}
        \centering
        \includegraphics[
            trim=0cm 0cm 1.5cm 0cm, 
            clip,
            width=\textwidth
        ]{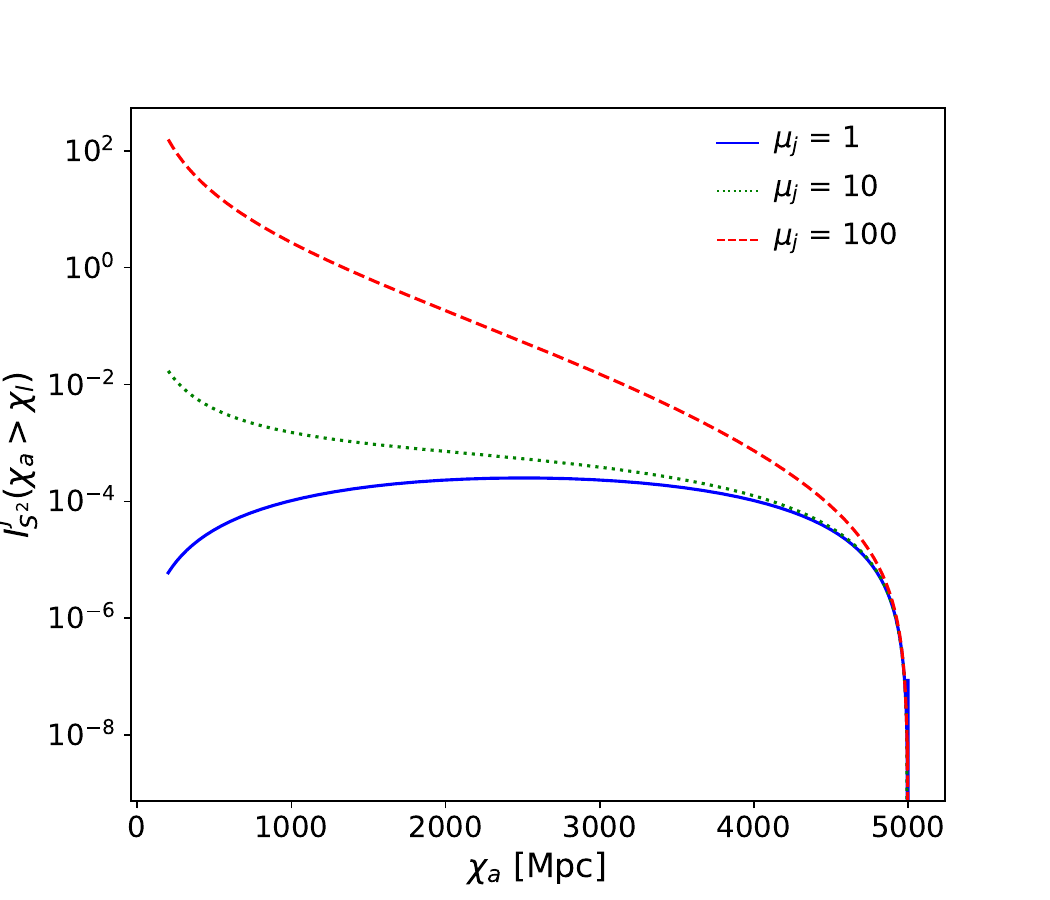}
    \end{subfigure}

    \caption{The $\chi_a$-dependence of the kernel functions $I^{j}_{S^{2}}(\chi_{a}<\chi_{l})$ (left panel) and $I^{j}_{S^{2}}(\chi_{a}>\chi_{l})$ (right panel) for $k_{\perp}=8\times10^5h\ \text{Mpc}^{-1}$. Here, to emphasize the shift in the value of $\chi_a$ at which the kernel function peaks, we assume a strong lens distance of $\chi_l = 200\ \text{Mpc}$, an observer distance of $\chi_s = 5 \times 10^3\ \text{Mpc}$, and a gravitational wave frequency of $10\ \text{Hz}$.}
    \label{fig:main2}
\end{figure}

where $I^{j}_{K^{2}}$, $I^{j}_{S^{2}}$, and $I^{j}_{KS}$ are defined as
\begin{align}
    I^{j}_{K^{2}}(\sigma)&\defeq\frac{1-\cos{(A^{j,+}(\sigma))J_{0}(A^{j,-}(\sigma))}}{2A^{2}},\\
    I^{j}_{S^{2}}(\sigma)&\defeq\frac{3-4\cos{(A^{j,+}(\sigma)/2)}J_{0}(A^{j,-}(\sigma)/2)+\cos(A^{j,+}(\sigma))J_{0}(A^{j,-}(\sigma))}{2A^{2}},\\
    I^{j}_{KS}(\sigma)&\defeq\frac{\sin{(A^{j,+}(\sigma))}J_{0}(A^{j,-}(\sigma))-2\sin{\qty(A^{j,+}(\sigma)/2)}J_{0}(A^{j,-}(\sigma)/2)}{2A^{2}},
\end{align}
where $J_{0}$ is the Bessel function, and $\sigma$ denotes the condition $\chi_{a}<\chi_{l}$ or $\chi_{a}>\chi_{l}$.
Similarly, $\langle K_{j}\rangle_{W}$ and $\langle S_{j}\rangle_{W}$ are obtained by substituting Eq. \eqref{pppk} into Eqs. \eqref{<k>} and \eqref{<s>}:
\begin{align}
    \langle K_{j}\rangle_{W}&=-\int^{\chi_{l}}_{0}d\chi_{a}\int\frac{k_{a\perp}d k_{a\perp}}{2\pi}P^{j}_{\kappa,\text{loc}}(\chi_{a},k_{a\perp})I^{j}_{K^{2}}(\chi_{a}<\chi_{l})-\int^{\chi_{s}}_{\chi_{l}}d\chi_{a}\int\frac{k_{a\perp} d k_{a\perp}}{2\pi}P^{j}_{\kappa,\text{loc}}(\chi_{a},k_{a\perp})I^{j}_{K^{2}}(\chi_{a}>\chi_{l}),\\
    \langle S_{j}\rangle_{W}&=\int^{\chi_{l}}_{0}d\chi_{a}\int\frac{k_{a\perp}d k_{a\perp}}{2\pi}P^{j}_{\kappa,\text{loc}}(\chi_{a},k_{a\perp})I^{j}_{S}(\chi_{a}<\chi_{l})+\int^{\chi_{s}}_{\chi_{l}}d\chi_{a}\int\frac{k_{a\perp} d k_{a\perp}}{2\pi}P^{j}_{\kappa,\text{loc}}(\chi_{a},k_{a\perp})I^{j}_{S}(\chi_{a}>\chi_{l}),
\end{align}

where $I^{j}_{S}$ is defined as follows:
\begin{align}
    I^{j}_{S}(\sigma)&\defeq\frac{A^{j,+}(\sigma)-\sin{(A^{j,+}(\sigma))}J_{0}(A^{j,-}(\sigma))}{2A^{2}}.\label{3.71}
\end{align}

\begin{figure}[t!]
    \centering 

    \begin{subfigure}[b]{0.495\textwidth}
        \centering
        \includegraphics[
            trim=0cm 0cm 1.5cm 0cm, 
            clip,
            width=\textwidth
        ]{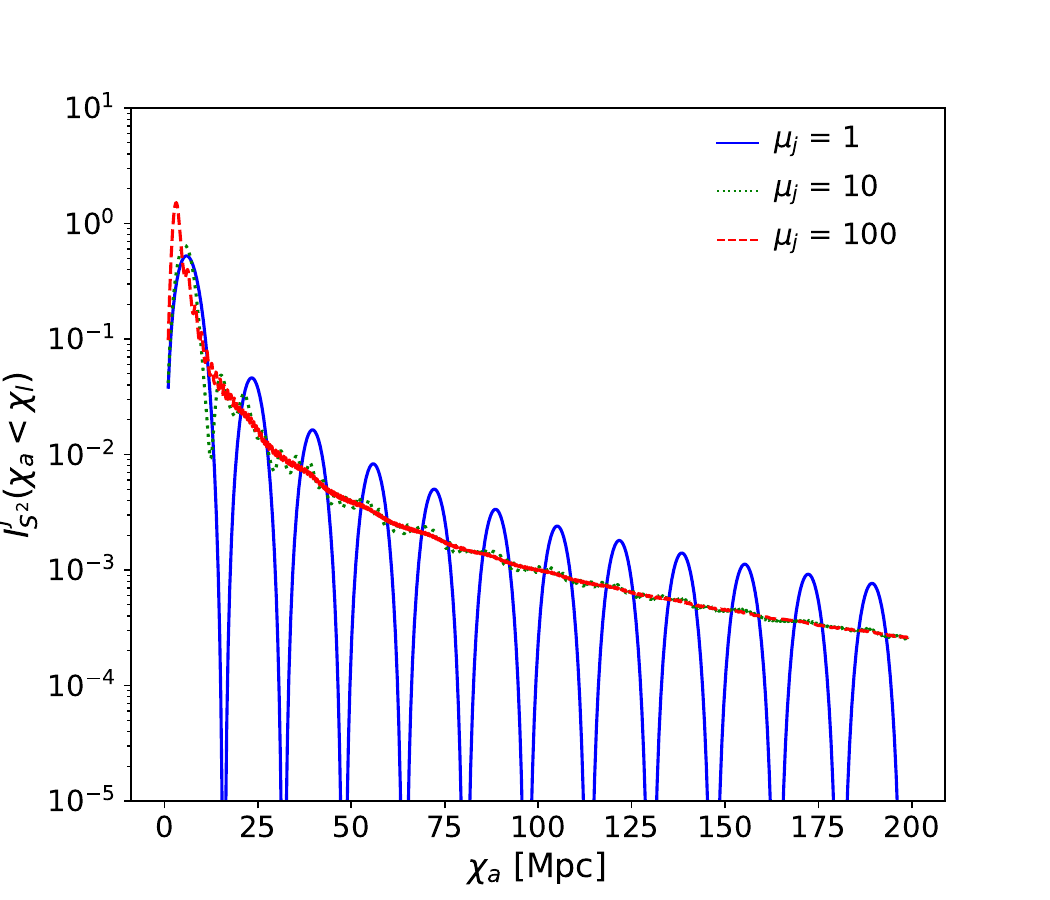}
    \end{subfigure}
    \hfill 
    \begin{subfigure}[b]{0.495\textwidth}
        \centering
        \includegraphics[
            trim=0cm 0cm 1.5cm 0cm, 
            clip,
            width=\textwidth
        ]{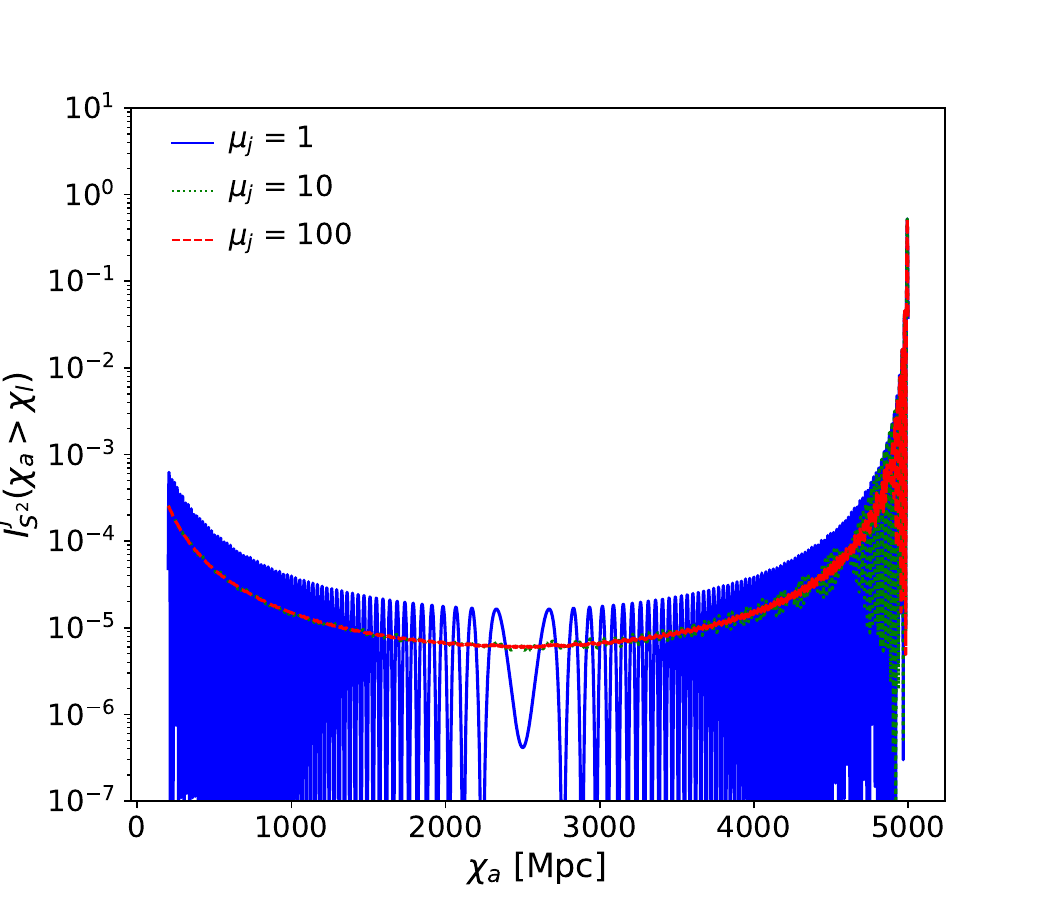}
    \end{subfigure}

    \caption{The $\chi_a$-dependence of the kernel functions $I^{j}_{S^{2}}(\chi_{a}<\chi_{l})$ (left panel) and $I^{j}_{S^{2}}(\chi_{a}>\chi_{l})$ (right panel) for $k_{\perp}=10^8h\ \text{Mpc}^{-1}$. Here, to emphasize the shift in the value of $\chi_a$ at which the kernel function peaks, we assume a strong lens distance of $\chi_l = 200\ \text{Mpc}$, an observer distance of $\chi_s = 5 \times 10^3\ \text{Mpc}$, and a gravitational wave frequency of $10\ \text{Hz}$.}
    \label{fig:main3}
\end{figure}

Figure \ref{fig:main} shows one of the kernel functions for $\langle S^{2}_{j}\rangle_{W}, I^{j}_{S^{2}}(\chi_{a}<\chi_{l})$ (left panel), and the other for $\langle S_{j}\rangle_{W}, I^{j}_{S}(\chi_{a}<\chi_{l})$ (right panel). For the case of $\mu_{j}=1$ (solid line), the graph peaks at $k\sim 1/r_{F}\sim 6\times10^{6} h\text{Mpc}^{-1}$, indicating that the influence of the matter power spectrum at the Fresnel scale is emphasized. As seen in the plot of $I^{j}_{S^{2}}(\chi_{a}<\chi_{l})$, an increase in the magnification factor $\mu_{j}$ shifts the effective Fresnel scale towards larger scales and enhances the amplitude of the kernel function. This result is similar to the findings of previous work \cite{oguri:strong}. Furthermore, the plot of $I^{j}_{S}(\chi_{a}<\chi_{l})$ indicates that the kernel function for the mean value, which arises from second-order terms of the weak lensing potential, exhibits a similar trend. We have also confirmed that the case of $\chi_{a}>\chi_{l}$ and other kernel functions, $I_{K^{2}}^{j}$ and $I_{KS}^{j}$, exhibit similar trends. 

Moreover, since we are considering cosmological density fluctuations, the impact of the magnification factor varies depending on the value of $\chi_a$. The shape of the kernel as a function of $\chi_a$ also depends on the value of $k_\perp$. As shown below, it exhibits distinct behaviors depending on whether $k_\perp$ is larger or smaller than the wavenumber corresponding to the peak scale (i.e., the effective Fresnel scale) of the kernel function in Fig. \ref{fig:main}. Figure \ref{fig:main2} shows the $\chi_a$-dependence of the kernel functions $I^{j}_{S^{2}}(\chi_{a}<\chi_{l})$ (left panel) and $I^{j}_{S^{2}}(\chi_{a}>\chi_{l})$ (right panel) for $k_{\perp}=8\times10^5h\ \text{Mpc}^{-1}$. This corresponds to considering a wavenumber regime smaller than that of the effective Fresnel scale for $\mu_j=100$. Note that we employ a different set of parameters from Fig. \ref{fig:main} to emphasize the shift in the value of $\chi_a$ at which the kernel function peaks. For $\mu_j=1$, the kernel function peaks around $\chi_a \simeq \chi_s/2 = 2.5 \times 10^3\ \text{Mpc}$, whereas for $\mu_j=10$ and $100$, it takes its maximum value at $\chi_a \simeq \chi_l = 200\ \text{Mpc}$. This result implies that the magnification effect due to the strong lens is particularly pronounced for weak lenses located in the immediate vicinity of the strong lens in this case. We have confirmed across all statistics at $k_{\perp}=8\times10^5h\ \text{Mpc}^{-1}$ that in the cases of $\mu_j =10$ and $100$, the fluctuation signal is most significantly enhanced when $\chi_a \simeq \chi_l$. Next, Fig. \ref{fig:main3} shows the kernel functions $I^{j}_{S^{2}}(\chi_{a}<\chi_{l})$ (left panel) and $I^{j}_{S^{2}}(\chi_{a}>\chi_{l})$ (right panel) for $k_{\perp}=10^8h\ \text{Mpc}^{-1}$. This corresponds to considering a wavenumber regime larger than that of the effective Fresnel scale for $\mu_j=1$. In this case, we find that the kernel functions are maximized near $\chi_a \simeq 0$ and $\chi_a \simeq \chi_s$ for all cases of $\mu_j=1, 10,$ and $100$. Similar behavior has been confirmed for the other statistics as well. These results indicate that the value of $\chi_a$ at which the kernel function is most magnified varies depending on the wavenumber.

\section{Conclusion}

In this paper, we have derived the amplification factor for gravitational waves, describing the modification in amplitude and phase due to strong gravitational lensing by galaxies or galaxy clusters and weak gravitational lensing by cosmological density fluctuations. We have then used these formulae to prove that the consistency relations of the same form as those in previous work \cite{mizuno:new} hold even in the presence of strong lensing, as considered in this study. These relations provide a crucial test to distinguish a true gravitational wave signal from non-physical artifacts, such as instrumental noise or systematic errors. Our findings confirm that the consistency relations serve as a robust test even for weak lensing signals induced by cosmological density fluctuations magnified by strong lensing.
Furthermore, we have investigated the dependence of these lensing statistics on the strong lensing magnification factor. Analysis of the kernel functions reveals that an increase in the magnification factor shifts the effective Fresnel scale towards larger scales and enhances the amplitude of the kernel function. This result indicates that strong lensing enhances the signal of cosmological density fluctuations on the effective Fresnel scale. We have also discussed the $\chi_a$-dependence of the kernel function in detail, and obtained its specific form as a function of $\chi_a$. Taken together, these results significantly raise the prospects for probing the small-scale structure of the universe using gravitational waves.

\section*{Acknowledgement}
T.S. acknowledges support from JSPS KAKENHI grant (Grant Number JP23K03411).

\begin{appendices}

\section{The Amplification Factor in the Path Integral Formalism} \label{apeA}

In this appendix, we derive Eqs. \eqref{F0f}, \eqref{F1f}, and \eqref{F2f} from the path integral representation of the amplification factor. Following Eq. \eqref{grapo}, we decompose the total gravitational potential into the strong lensing potential $\psi$ and the weak lensing potential $\Phi_{W}$, and the path integral representation of the amplification factor for the scalar wave is given by \cite{Nakamura}
\begin{align}
    F(\omega,\chi_{s})&=\int \mathscr{D}\bm{\theta}\exp{i\omega\int^{\chi_{s}}_{0} d\chi\left[ \frac{1}{2}\chi^{2}|\dot{\bm{\theta}}|^{2}-\psi(\bm{\theta}_{l})\delta(\chi-\chi_{l})-2\Phi_{W}(\chi,\bm{\theta})\right]}. \label{1.1}
\end{align}
where the dot denotes the derivative with respect to $\chi$, and the measure of the path integral is normalized such that $F=1$ when $\psi=\Phi_{W}=0$.
Since $\Phi_{W}$ can be considered small, Taylor expanding only the part concerning $\Phi_{W}$ gives
\begin{align}
    F(\omega,\chi_{s})&=\int \mathscr{D}\bm{\theta}\exp{i\omega\int^{\chi_{s}}_{0} d\chi\left[ {\frac{1}{2}\chi^{2}|\dot{\bm{\theta}}|^{2}-\psi(\bm{\theta}_{l})\delta(\chi-\chi_{l})}\right]}\nonumber\\
    &\times\Bigg\{1-2i\omega\int^{\chi_{s}}_{0}d\chi_{a} \Phi_{W}(\chi_{a},\bm{\theta}_{a})+(2i\omega)^{2}\int^{\chi_{s}}_{0}d\chi_{b}\int^{\chi_{b}}_{0}d\chi_{a}\Phi_{W}(\chi_{b},\bm{\theta}_{b})\Phi_{W}(\chi_{a},\bm{\theta}_{a})\nonumber\\ 
    &+\mathcal{O}(\Phi_{W}^{3}) \Bigg\}, \label{1.2}
\end{align}
where we have used the symmetry between $\chi_{a}$ and $\chi_{b}$ to combine the term with $\chi_{a}>\chi_{b}$ into that with $\chi_{a}<\chi_{b}$.
To evaluate the path integral through discretization, we use the following relations \cite{Nakamura,oguri:wave}
\begin{align}
    &\epsilon\defeq\chi_{k+1}-\chi_{k},\quad \tau_{ab}\defeq\frac{\chi_{a}\chi_{b}}{\chi_{b}-\chi_{a}},\quad A_{k}\defeq\frac{2\pi i}{\omega}\frac{1}{\tau_{k(k+1)}}.
\end{align}
The zeroth-order term of Eq. \eqref{1.2} in $\Phi_{W}$ is derived as
\begin{align}
    F_{0}&=\left[\prod_{k=1}^{N}\int\frac{d\bm{\theta}_{k}}{A_{k}}\right]\exp{i\omega\epsilon\sum_{k=1}^{N}\frac{\chi_{k}\chi_{k+1}}{2}\left|\frac{\bm{\theta}_{k+1}-\bm{\theta}_{k}}{\epsilon}\right|^{2}-i\omega\psi(\bm{\theta}_{l})}\nonumber\\
    &=\frac{\omega
    }{2\pi i}\tau_{ls}\int d\bm{\theta}_{l}\exp{i\omega \tau_{ls}\frac{|\bm{\theta}_{l}-\bm{\theta}_{s}|^{2}}{2}-i\omega\psi(\bm{\theta}_{l})}. \label{1.6}
\end{align}
where $s=N+1$, and we have used the following relation \cite{Nakamura}:
\begin{align}
    \sum_{k=l}^{N}\tau_{k(k+1)}|\bm{\theta}_{k}-\bm{\theta}_{k+1}|^{2}=\tau_{ls}|\bm{\theta}_{l}-\bm{\theta}_{s}|^{2}+\sum_{k=l+1}^{N}\frac{\tau_{k(k+1)}\tau_{lk}}{\tau_{l(k+1)}}\left|\bm{\theta}_{k}-\frac{\tau_{l(k+1)}}{\tau_{k(k+1)}}\bm{\theta}_{l}-\frac{\tau_{l(k+1)}}{\tau_{lk}}\bm{\theta}_{k+1}\right|^{2}, \label{1.11}
\end{align}

Next, we calculate the first-order term of Eq. \eqref{1.2} in $\Phi_{W}$. It is described by
\begin{align}
    F_{1}&=-2i\omega\left[\prod_{k=1}^{N}\int\frac{d\bm{\theta}_{k}}{A_{k}}\right]\exp{i\omega\epsilon\sum_{k=1}^{N}\frac{\chi_{k}\chi_{k+1}}{2}\left|\frac{\bm{\theta}_{k+1}-\bm{\theta}_{k}}{\epsilon}\right|^{2}-i\omega\psi(\bm{\theta}_{l})}\times \epsilon\sum_{a=1}^{N}\Phi_{W}(\chi_{a},\bm{\theta}_{a}).
\end{align}
The calculation differs for the cases $a<l$ and $a>l$, so we evaluate each case separately. For $a<l$, the integral yields:
\begin{align}  
     F_{1(a<l)}=-2i\omega\epsilon\sum^{l-1}_{a=1}\int\frac{d\bm{\theta}_{a}}{A_{a}}\Phi_{W}(\chi_{a},\bm{\theta}_{a})\left[\prod_{k=a+1}^{N}\int\frac{d\bm{\theta}_{k}}{A_{k}}\right]\exp{i\omega\sum_{k=a}^{N}\tau_{k(k+1)}\frac{|\bm{\theta}_{k}-\bm{\theta}_{k+1}|^{2}}{2}-i\omega\psi(\bm{\theta}_{l})}. \label{1.10}
\end{align}
Using Eq. \eqref{1.11}, we calculate Eq. \eqref{1.10} as follows:
\begin{align}
    F_{1(a<l)}&= -2i\omega\epsilon\sum^{l-1}_{a=1}\int\frac{d\bm{\theta}_{a}}{A_{a}}\Phi_{W}(\chi_{a},\bm{\theta}_{a})\exp{\frac{i\omega}{2}\tau_{as}|\bm{\theta}_{a}-\bm{\theta}_{s}|^{2}}\frac{\tau_{al}}{\tau_{a(a+1)}}
     \left[\prod_{k=l}^{N}\int\frac{d\bm{\theta}_{k}}{A_{k}}\right]\nonumber\\
      &\times\exp{i\omega\sum_{k=l}^{N}\frac{\tau_{k(k+1)}\tau_{ak}}{2\tau_{a(k+1)}}\left|\bm{\theta}_{k}-\frac{\tau_{a(k+1)}}{\tau_{k(k+1)}}\bm{\theta}_{a}-\frac{\tau_{a(k+1)}}{\tau_{ak}}\bm{\theta}_{k+1}\right|^{2}-i\omega\psi(\bm{\theta}_{l})}. \label{1.12}
\end{align}
We find a new relation:
\begin{align}
    &\sum^{N}_{k=l}\frac{\tau_{k(k+1)}\tau_{ak}}{2\tau_{a(k+1)}}\qty|\bm{\theta}_{k}-\frac{\tau_{a(k+1)}}{\tau_{k(k+1)}}\bm{\theta}_{a}-\frac{\tau_{a(k+1)}}{\tau_{ak}}\bm{\theta}_{k+1}|^{2}\nonumber\\ &=\tau_{ls}\frac{|\bm{\theta}_{l}-\bm{\theta}_{s}|^{2}}{2}-\tau_{as}\frac{|\bm{\theta}_{a}-\bm{\theta}_{s}|^{2}}{2}+\tau_{al}\frac{|\bm{\theta}_{l}-\bm{\theta}_{a}|^{2}}{2}+\sum^{N}_{k=l+1}\frac{\tau_{k(k+1)}\tau_{lk}}{2\tau_{l(k+1)}}\qty|\bm{\theta}_{k}-\frac{\tau_{l(k+1)}}{\tau_{k(k+1)}}\bm{\theta}_{l}-\frac{\tau_{l(k+1)}}{\tau_{lk}}\bm{\theta}_{k+1}|^{2}. \label{1.13}
\end{align}
Substituting Eq. \eqref{1.13} into Eq. \eqref{1.12} yields:
\begin{align}
    F_{1(a<l)}&=-2i\omega\qty(\frac{\omega}{2\pi i})^{2}\epsilon\sum^{l-1}_{a=1}\tau_{al}\tau_{ls}\int d\bm{\theta}_{a}\Phi_{W}(\chi_{a},\bm{\theta}_{a})\int d\bm{\theta}_{l}\exp{i\omega\qty[\tau_{ls}\frac{|\bm{\theta}_{l}-\bm{\theta}_{s}|^{2}}{2}+\tau_{al}\frac{|\bm{\theta}_{l}-\bm{\theta}_{a}|^{2}}{2}-\psi(\bm{\theta}_{l})]}. \label{1.14}
\end{align}
Restoring the discrete variable to the continuous variable, Eq. \eqref{1.14} can be rewritten as follows:
\begin{align}
    F_{1(a<l)}&=\int^{\chi_{l}}_{0}d\chi_{a}\int d\bm{\theta}_{a}\int d\bm{\theta}_{l}\frac{2\omega}{i}\Phi_{W}(\chi_{a},\bm{\theta}_{a})\qty(\frac{\omega}{2\pi i})^{2}\tau_{al}\tau_{ls}\exp{i\omega\qty[\tau_{al}\frac{|\bm{\theta}_{a}-\bm{\theta}_{l}|^{2}}{2}+\tau_{ls}\frac{|\bm{\theta}_{l}-\bm{\theta}_{s}|^{2}}{2}-\psi(\bm{\theta}_{l})]}. \label{1.15}
\end{align}

Next, for the case $a>l$, we calculate the first-order term in $\Phi_{W}$. The integral yields:
\begin{align}
    F_{1(a>l)}&=-2i\omega\epsilon\sum_{a=l+1}^{N}\left[\prod_{k=1}^{N}\int\frac{d\bm{\theta}_{k}}{A_{k}}\right]\Phi_{W}(\chi_{a},\bm{\theta}_{a})\exp{i\omega\sum_{k=1}^{N}\tau_{k(k+1)}\frac{|\bm{\theta}_{k}-\bm{\theta}_{k+1}|^{2}}{2}-i\omega\psi(\bm{\theta}_{l})}.
\end{align}
Following the same procedure used to derive Eqs. \eqref{1.10} to \eqref{1.15}, we obtain:
\begin{align}
    F_{1(a>l)}&=\int^{\chi_{s}}_{\chi_{l}}d\chi_{a}\int d\bm{\theta}_{a}\int d\bm{\theta}_{l}\frac{2\omega}{i}\Phi_{W}(\chi_{a},\bm{\theta}_{a})\qty(\frac{\omega}{2\pi i})^{2}\tau_{la}\tau_{as}\exp{i\omega\qty[\tau_{la}\frac{|\bm{\theta}_{l}-\bm{\theta}_{a}|^{2}}{2}+\tau_{as}\frac{|\bm{\theta}_{a}-\bm{\theta}_{s}|^{2}}{2}-\psi(\bm{\theta}_{l})]}. \label{1.22}
\end{align}
The above equation is simply Eq. \eqref{1.15} with $k$ and $l$ interchanged. $F_{1}$ is given by the sum of Eqs. \eqref{1.15} and \eqref{1.22} as $F_{1}=F_{1(a<l)}+F_{1(a>l)}$, which is indeed consistent with Eq. \eqref{F1}.

Next, we evaluate the second-order corrections to the amplification factor. For the configuration where both weak lenses are situated between the strong lens and the source $(\chi_{a}<\chi_{b}<\chi_{l})$, the second-order term in $\Phi_{W}$ from Eq. \eqref{1.2} is initially expressed as
\begin{align}
    F_{2(a,b<l)}&=-4\omega^{2}\sum^{l-1}_{b=1}\epsilon\sum^{b-1}_{a=1}\epsilon\qty[\prod^{N}_{k=1}\int\frac{d\bm{\theta}_{k}}{A_{k}}]\Phi_{W}(\chi_{a},\bm{\theta}_{a})\Phi_{W}(\chi_{b},\bm{\theta}_{b})
    \exp{i\omega\sum_{k=1}^{N}\tau_{k(k+1)}\frac{|\bm{\theta}_{k}-\bm{\theta}_{k+1}|^{2}}{2}-i\omega\psi(\bm{\theta}_{l})}.
\end{align}
Following a procedure similar to the one used to derive $F_{1}$ and by repeatedly applying Eq. \eqref{1.13}, this is eventually reduced to the following simplified form:
\begin{align}
    F_{2(a,b<l)}&=\int^{\chi_{l}}_{0}d\chi_{b}\int^{\chi_{b}}_{0}d\chi_{a}\int d\bm{\theta}_{a}\int d\bm{\theta}_{b}\int d\bm{\theta}_{l}\frac{2\omega}{i}\Phi_{W}(\chi_{a},\bm{\theta}_{a})\frac{2\omega}{i}\Phi_{W}(\chi_{b},\bm{\theta}_{b})\nonumber\\
    &\times \qty(\frac{\omega}{2\pi i})^{3}\tau_{ab}\tau_{bl}\tau_{ls}\exp\Bigg\{i\omega\qty(\tau_{ab}\frac{|\bm{\theta}_{a}-\bm{\theta}_{b}|^{2}}{2}+\tau_{bl}\frac{|\bm{\theta}_{b}-\bm{\theta}_{l}|^{2}}{2}+\tau_{ls}\frac{|\bm{\theta}_{l}-\bm{\theta}_{s}|^{2}}{2}-\psi(\bm{\theta}_{l}))\Bigg\}. \label{1.34}
\end{align}

The other cases, $(\chi_{l}<\chi_{a}<\chi_{b})$ and $(\chi_{a}<\chi_{l}<\chi_{b})$, are derived through the same procedure and are given as follows, respectively:
\begin{align}
    F_{2(a,b>l)}&=\int^{\chi_{s}}_{\chi_{l}}d\chi_{b}\int^{\chi_{b}}_{\chi_{l}}d\chi_{a}\int d\bm{\theta}_{a}\int d\bm{\theta}_{b}\int d\bm{\theta}_{l}\frac{2\omega}{i}\Phi_{W}(\chi_{a},\bm{\theta}_{a})\frac{2\omega}{i}\Phi_{W}(\chi_{b},\bm{\theta}_{b})\nonumber\\
    &\times \qty(\frac{\omega}{2\pi i})^{3}\tau_{la}\tau_{ab}\tau_{bs}\exp\Bigg\{i\omega\qty(\tau_{la}\frac{|\bm{\theta}_{l}-\bm{\theta}_{a}|^{2}}{2}+\tau_{ab}\frac{|\bm{\theta}_{a}-\bm{\theta}_{b}|^{2}}{2}+\tau_{bs}\frac{|\bm{\theta}_{b}-\bm{\theta}_{s}|^{2}}{2}-\psi(\bm{\theta}_{l}))\Bigg\}, \label{1.41}
    \\
    F_{2(a<l,b>l)}&=\int^{\chi_{s}}_{\chi_{l}}d\chi_{b}\int^{\chi_{l}}_{0}d\chi_{a}\int d\bm{\theta}_{a}\int d\bm{\theta}_{b}\int d\bm{\theta}_{l}\frac{2\omega}{i}\Phi_{W}(\chi_{a},\bm{\theta}_{a})\frac{2\omega}{i}\Phi_{W}(\chi_{b},\bm{\theta}_{b})\nonumber\\
    &\times \qty(\frac{\omega}{2\pi i})^{3}\tau_{al}\tau_{lb}\tau_{bs}\exp\Bigg\{i\omega\qty(\tau_{al}\frac{|\bm{\theta}_{a}-\bm{\theta}_{l}|^{2}}{2}+\tau_{lb}\frac{|\bm{\theta}_{l}-\bm{\theta}_{b}|^{2}}{2}+\tau_{bs}\frac{|\bm{\theta}_{b}-\bm{\theta}_{s}|^{2}}{2}-\psi(\bm{\theta}_{l}))\Bigg\}. \label{1.41b}
\end{align}
$F_{2}$ is given by the sum of Eqs. \eqref{1.34}, \eqref{1.41} and \eqref{1.41b} as $F_{2}=F_{2(a,b<l)}+F_{2(a,b>l)}+F_{2(a<l,b>l)}$, which is indeed consistent with Eq. \eqref{F2}.

\section{Explicit Expressions for the amplitude and phase fluctuations and the Treatment of Delta Functions} \label{apeB}

As stated in Subsection 3.1, utilizing Eqs. \eqref{F0f}-\eqref{F2f} and \eqref{ks1}-\eqref{shapi} yields the explicit expressions for $K_{j1}$, $K_{j2}$, $S_{j1}$, and $S_{j2}$. These results are summarized below:
\begin{align}
    K_{j1}&=\int^{\chi_{l}}_{0}d\chi_{a}\int\frac{d\bm{k}_{a}}{(2\pi)^{3}}(-2\omega)\Tilde{\Phi}_{W}(\chi_{a},\bm{k}_{a})e^{ik_{a\parallel} \chi_{a}+i\bm{k}_{a\perp}\cdot\chi_{a}\bm{\theta}_{l,j}}\nonumber\\ 
    &\times\sin{\Bigg\{\frac{(\chi_{a} k_{a\perp1})^{2}}{2\omega}\qty(\frac{1}{\tau_{al}}+\frac{\mu_{j,1}}{\tau_{ls}})+\frac{(\chi_{a} k_{a\perp2})^{2}}{2\omega}\qty(\frac{1}{\tau_{al}}+\frac{\mu_{j,2}}{\tau_{ls}})\Bigg\}}\nonumber\\
    &+\int^{\chi_{s}}_{\chi_{l}}d \chi_{a}\int\frac{d\bm{k}_{a}}{(2\pi)^{3}}(-2\omega)\Tilde{\Phi}(\chi_{a},\bm{k}_{a})e^{ik_{a\parallel}\chi_{a} +i\bm{k}_{a\perp}\cdot\chi_{a}\qty(\frac{\tau_{la}\bm{\theta}_{l,j}+\tau_{as}\bm{\theta}_{s}}{\tau_{la}+\tau_{as}})}\nonumber\\
    &\times\sin{\Bigg\{\frac{(\chi_{a}k_{a\perp1})^{2}}{2\omega}\frac{\tau_{ls}}{\tau_{as}}\qty(\frac{1}{\tau_{la}}+\frac{\mu_{j,1}}{\tau_{as}})+\frac{(\chi_{a}k_{a\perp2})^{2}}{2\omega}\frac{\tau_{ls}}{\tau_{as}}\qty(\frac{1}{\tau_{la}}+\frac{\mu_{j,2}}{\tau_{as}})\Bigg\}},\label{kj1spe}
    \\
    S_{j1}&=\int^{\chi_{l}}_{0}d\chi_{a}\int\frac{d\bm{k}_{a}}{(2\pi)^{3}}(-2\omega)\Tilde{\Phi}_{W}(\chi_{a},\bm{k}_{a})e^{ik_{a\parallel} \chi_{a}+i\bm{k}_{a\perp}\cdot\chi_{a}\bm{\theta}_{l,j}}\nonumber\\ 
    &\times\Bigg[\cos{\Bigg\{\frac{(\chi_{a} k_{a\perp1})^{2}}{2\omega}\qty(\frac{1}{\tau_{al}}+\frac{\mu_{j,1}}{\tau_{ls}})+\frac{(\chi_{a} k_{a\perp2})^{2}}{2\omega}\qty(\frac{1}{\tau_{al}}+\frac{\mu_{j,2}}{\tau_{ls}})\Bigg\}}-1\Bigg]\nonumber\\
    &+\int^{\chi_{s}}_{\chi_{l}}d \chi_{a}\int\frac{d\bm{k}_{a}}{(2\pi)^{3}}(-2\omega)\Tilde{\Phi}(\chi_{a},\bm{k}_{a})e^{ik_{a\parallel}\chi_{a} +i\bm{k}_{a\perp}\cdot\chi_{a}\qty(\frac{\tau_{la}\bm{\theta}_{l,j}+\tau_{as}\bm{\theta}_{s}}{\tau_{la}+\tau_{as}})}\nonumber\\
    &\times\Bigg[\cos{\Bigg\{\frac{(\chi_{a}k_{a\perp1})^{2}}{2\omega}\frac{\tau_{ls}}{\tau_{as}}\qty(\frac{1}{\tau_{la}}+\frac{\mu_{j,1}}{\tau_{as}})+\frac{(\chi_{a}k_{a\perp2})^{2}}{2\omega}\frac{\tau_{ls}}{\tau_{as}}\qty(\frac{1}{\tau_{la}}+\frac{\mu_{j,2}}{\tau_{as}})\Bigg\}}-1\Bigg],\label{sj1spe}
\end{align}
\begin{align}  
    K_{j2}&=\int^{\chi_{l}}_{0}d\chi_{b}\int^{\chi_{b}}_{0}d\chi_{a}\int\frac{d\bm{k}_{a}}{(2\pi)^{3}}\int\frac{d\bm{k}_{b}}{(2\pi)^{3}}(-4\omega^{2})\Tilde{\Phi}_{W}(\chi_{a},\bm{k}_{a})\Tilde{\Phi}_{W}(\chi_{b},\bm{k}_{b})e^{ik_{a\parallel}\chi_{a}+ik_{b\parallel}\chi_{b}+i(\bm{k}_{a\perp}\chi_{a}+\bm{k}_{b\perp}\chi_{b})\cdot\bm{\theta}_{l,j}}\nonumber\\
    &\times\Bigg[\cos{\Bigg\{\frac{(\chi_{a}\bm{k}_{a\perp})^{2}}{2\omega\tau_{ab}}+\frac{(\chi_{a}k_{a\perp1}+\chi_{b}k_{b\perp1})^{2}}{2\omega}\qty(\frac{1}{\tau_{bl}}+\frac{\mu_{j,1}}{\tau_{ls}})+\frac{(\chi_{a}k_{a\perp2}+\chi_{b}k_{b\perp2})^{2}}{2\omega}\qty(\frac{1}{\tau_{bl}}+\frac{\mu_{j,2}}{\tau_{ls}})\Bigg\}}\nonumber\\
    &-\cos{\Bigg\{\frac{(\chi_{a} k_{a\perp1})^{2}}{2\omega}\qty(\frac{1}{\tau_{al}}+\frac{\mu_{j,1}}{\tau_{ls}})+\frac{(\chi_{a} k_{a\perp2})^{2}}{2\omega}\qty(\frac{1}{\tau_{al}}+\frac{\mu_{j,2}}{\tau_{ls}})+\frac{(\chi_{b} k_{b\perp1})^{2}}{2\omega}\qty(\frac{1}{\tau_{bl}}+\frac{\mu_{j,1}}{\tau_{ls}})+\frac{(\chi_{b} k_{b\perp2})^{2}}{2\omega}\qty(\frac{1}{\tau_{bl}}+\frac{\mu_{j,2}}{\tau_{ls}})\Bigg\}}\Bigg]\nonumber\\
    &+\int^{\chi_{s}}_{\chi_{l}}d\chi_{b}\int^{\chi_{b}}_{\chi_{l}}d\chi_{a}\int\frac{d\bm{k}_{a}}{(2\pi)^{3}}\int\frac{d\bm{k}_{b}}{(2\pi)^{3}}(-4\omega^{2})\Tilde{\Phi}_{W}(\chi_{a},\bm{k}_{a})\Tilde{\Phi}_{W}(\chi_{b},\bm{k}_{b})e^{ik_{a\parallel}\chi_{a}+ik_{b\parallel}\chi_{b}}\nonumber\\
    &\times\exp{i\chi_{a}\bm{k}_{a\perp}\cdot\frac{\tau_{la}\bm{\theta}_{l,j}+\tau_{as}\bm{\theta}_{s}}{\tau_{la}+\tau_{as}}+i\chi_{b}\bm{k}_{b\perp}\cdot\frac{\tau_{lb}\bm{\theta}_{l,j}+\tau_{bs}\bm{\theta}_{s}}{\tau_{lb}+\tau_{bs}}}\nonumber\\
    &\times\Bigg[\cos\Bigg\{\frac{(\chi_{a}\bm{k}_{a\perp})^{2}}{2\omega(\tau_{la}+\tau_{ab})}+\frac{1}{2\omega(\tau_{lb}+\tau_{bs})}\qty(\chi_{b}\bm{k}_{b\perp}+\frac{\tau_{lb}}{\tau_{la}}\chi_{a}\bm{k}_{a\perp})^{2}\nonumber
    \\
    &+\frac{\mu_{j,1}}{2\omega\tau_{ls}}\qty(\frac{\tau_{ls}}{\tau_{as}}\chi_{a}k_{a\perp1}+\frac{\tau_{ls}}{\tau_{bs}}\chi_{b}k_{b\perp1})^{2}+\frac{\mu_{j,2}}{2\omega\tau_{ls}}\qty(\frac{\tau_{ls}}{\tau_{as}}\chi_{a}k_{a\perp2}+\frac{\tau_{ls}}{\tau_{bs}}\chi_{b}k_{b\perp2})^{2}\Bigg\}\nonumber\\
    &-\cos\Bigg\{\frac{(\chi_{a}k_{a\perp1})^{2}}{2\omega}\frac{\tau_{ls}}{\tau_{as}}\qty(\frac{1}{\tau_{la}}+\frac{\mu_{j,1}}{\tau_{as}})+\frac{(\chi_{a}k_{a\perp2})^{2}}{2\omega}\frac{\tau_{ls}}{\tau_{as}}\qty(\frac{1}{\tau_{la}}+\frac{\mu_{j,2}}{\tau_{as}})\nonumber\\
    &+\frac{(\chi_{b}k_{b\perp1})^{2}}{2\omega}\frac{\tau_{ls}}{\tau_{bs}}\qty(\frac{1}{\tau_{lb}}+\frac{\mu_{j,1}}{\tau_{bs}})+\frac{(\chi_{b}k_{b\perp2})^{2}}{2\omega}\frac{\tau_{ls}}{\tau_{bs}}\qty(\frac{1}{\tau_{lb}}+\frac{\mu_{j,2}}{\tau_{bs}})\Bigg\}\Bigg]\nonumber
\end{align}
\begin{align}
    &+\int^{\chi_{s}}_{\chi_{l}}d\chi_{b}\int^{\chi_{l}}_{0}d\chi_{a}\int\frac{d\bm{k}_{a}}{(2\pi)^{3}}\int\frac{d\bm{k}_{b}}{(2\pi)^{3}}(-4\omega^{2})\Tilde{\Phi}_{W}(\chi_{a},\bm{k}_{a})\Tilde{\Phi}_{W}(\chi_{b},\bm{k}_{b})e^{ik_{a\parallel}\chi_{a}+ik_{b\parallel}\chi_{b}}
    \nonumber\\
    &\times\exp{i\chi_{a}\bm{k}_{a\perp}\cdot\bm{\theta}_{l,j}+i\chi_{b}\bm{k}_{b\perp}\cdot\frac{\tau_{lb}\bm{\theta}_{l,j}+\tau_{bs}\bm{\theta}_{s}}{\tau_{lb}+\tau_{bs}}}\nonumber\\
    &\times\Bigg[\cos\Bigg\{\frac{(\chi_{a}\bm{k}_{a\perp})^{2}}{2\omega\tau_{al}}+\frac{(\chi_{b}\bm{k}_{b\perp})^{2}}{2\omega(\tau_{lb}+\tau_{bs})}+\frac{\mu_{j,1}}{2\omega\tau_{ls}}\qty(\chi_{a}k_{a\perp1}+\frac{\tau_{ls}}{\tau_{bs}}\chi_{b}k_{b\perp1})^{2}+\frac{\mu_{j,2}}{2\omega\tau_{ls}}\qty(\chi_{a}k_{a\perp2}+\frac{\tau_{ls}}{\tau_{bs}}\chi_{b}k_{b\perp2})^{2}\Bigg\}\nonumber\\
    &-\cos\Bigg\{\frac{(\chi_{a} k_{a\perp1})^{2}}{2\omega}\qty(\frac{1}{\tau_{al}}+\frac{\mu_{j,1}}{\tau_{ls}})+\frac{(\chi_{a} k_{a\perp2})^{2}}{2\omega}\qty(\frac{1}{\tau_{al}}+\frac{\mu_{j,2}}{\tau_{ls}})\nonumber\\
    &+\frac{(\chi_{b}k_{b\perp1})^{2}}{2\omega}\frac{\tau_{ls}}{\tau_{bs}}\qty(\frac{1}{\tau_{lb}}+\frac{\mu_{j,1}}{\tau_{bs}})+\frac{(\chi_{b}k_{b\perp2})^{2}}{2\omega}\frac{\tau_{ls}}{\tau_{bs}}\qty(\frac{1}{\tau_{lb}}+\frac{\mu_{j,2}}{\tau_{bs}})\Bigg\}\Bigg],\label{kj2spe}
\end{align}
\begin{align}
    S_{j2}&=\int^{\chi_{l}}_{0}d\chi_{b}\int^{\chi_{b}}_{0}d\chi_{a}\int\frac{d\bm{k}_{a}}{(2\pi)^{3}}\int\frac{d\bm{k}_{b}}{(2\pi)^{3}}(-4\omega^{2})\Tilde{\Phi}_{W}(\chi_{a},\bm{k}_{a})\Tilde{\Phi}_{W}(\chi_{b},\bm{k}_{b})e^{ik_{a\parallel}\chi_{a}+ik_{b\parallel}\chi_{b}+i(\bm{k}_{a\perp}\chi_{a}+\bm{k}_{b\perp}\chi_{b})\cdot\bm{\theta}_{l,j}}\nonumber\\
    &\times\Bigg[-\sin{\Bigg\{\frac{(\chi_{a}\bm{k}_{a\perp})^{2}}{2\omega\tau_{ab}}+\frac{(\chi_{a}k_{a\perp1}+\chi_{b}k_{b\perp1})^{2}}{2\omega}\qty(\frac{1}{\tau_{bl}}+\frac{\mu_{j,1}}{\tau_{ls}})+\frac{(\chi_{a}k_{a\perp2}+\chi_{b}k_{b\perp2})^{2}}{2\omega}\qty(\frac{1}{\tau_{bl}}+\frac{\mu_{j,2}}{\tau_{ls}})\Bigg\}}\nonumber\\
    &+\sin{\Bigg\{\frac{(\chi_{a} k_{a\perp1})^{2}}{2\omega}\qty(\frac{1}{\tau_{al}}+\frac{\mu_{j,1}}{\tau_{ls}})+\frac{(\chi_{a} k_{a\perp2})^{2}}{2\omega}\qty(\frac{1}{\tau_{al}}+\frac{\mu_{j,2}}{\tau_{ls}})+\frac{(\chi_{b} k_{b\perp1})^{2}}{2\omega}\qty(\frac{1}{\tau_{bl}}+\frac{\mu_{j,1}}{\tau_{ls}})+\frac{(\chi_{b} k_{b\perp2})^{2}}{2\omega}\qty(\frac{1}{\tau_{bl}}+\frac{\mu_{j,2}}{\tau_{ls}})\Bigg\}}\nonumber\\
    &+\frac{\chi_{a}k_{a\perp1}\chi_{b}k_{b\perp1}}{\omega}\qty(\frac{1}{\tau_{bl}}+\frac{\mu_{j,1}}{\tau_{ls}})+\frac{\chi_{a}k_{a\perp2}\chi_{b}k_{b\perp2}}{\omega}\qty(\frac{1}{\tau_{bl}}+\frac{\mu_{j,2}}{\tau_{ls}})\Bigg]\nonumber\\
    &+\int^{\chi_{s}}_{\chi_{l}}d\chi_{b}\int^{\chi_{b}}_{\chi_{l}}d\chi_{a}\int\frac{d\bm{k}_{a}}{(2\pi)^{3}}\int\frac{d\bm{k}_{b}}{(2\pi)^{3}}(-4\omega^{2})\Tilde{\Phi}_{W}(\chi_{a},\bm{k}_{a})\Tilde{\Phi}_{W}(\chi_{b},\bm{k}_{b})e^{ik_{a\parallel}\chi_{a}+ik_{b\parallel}\chi_{b}}\nonumber\\
    &\times\exp{i\chi_{a}\bm{k}_{a\perp}\cdot\frac{\tau_{la}\bm{\theta}_{l,j}+\tau_{as}\bm{\theta}_{s}}{\tau_{la}+\tau_{as}}+i\chi_{b}\bm{k}_{b\perp}\cdot\frac{\tau_{lb}\bm{\theta}_{l,j}+\tau_{bs}\bm{\theta}_{s}}{\tau_{lb}+\tau_{bs}}}\nonumber\\
    &\times\Bigg[-\sin\Bigg\{\frac{(\chi_{a}\bm{k}_{a\perp})^{2}}{2\omega(\tau_{la}+\tau_{ab})}+\frac{1}{2\omega(\tau_{lb}+\tau_{bs})}\qty(\chi_{b}\bm{k}_{b\perp}+\frac{\tau_{lb}}{\tau_{la}}\chi_{a}\bm{k}_{a\perp})^{2}\nonumber
    \\
    &+\frac{\mu_{j,1}}{2\omega\tau_{ls}}\qty(\frac{\tau_{ls}}{\tau_{as}}\chi_{a}k_{a\perp1}+\frac{\tau_{ls}}{\tau_{bs}}\chi_{b}k_{b\perp1})^{2}+\frac{\mu_{j,2}}{2\omega\tau_{ls}}\qty(\frac{\tau_{ls}}{\tau_{as}}\chi_{a}k_{a\perp2}+\frac{\tau_{ls}}{\tau_{bs}}\chi_{b}k_{b\perp2})^{2}\Bigg\}\nonumber\\
    &+\sin\Bigg\{\frac{(\chi_{a}k_{a\perp1})^{2}}{2\omega}\frac{\tau_{ls}}{\tau_{as}}\qty(\frac{1}{\tau_{la}}+\frac{\mu_{j,1}}{\tau_{as}})+\frac{(\chi_{a}k_{a\perp2})^{2}}{2\omega}\frac{\tau_{ls}}{\tau_{as}}\qty(\frac{1}{\tau_{la}}+\frac{\mu_{j,2}}{\tau_{as}})\nonumber\\
    &+\frac{(\chi_{b}k_{b\perp1})^{2}}{2\omega}\frac{\tau_{ls}}{\tau_{bs}}\qty(\frac{1}{\tau_{lb}}+\frac{\mu_{j,1}}{\tau_{bs}})+\frac{(\chi_{b}k_{b\perp2})^{2}}{2\omega}\frac{\tau_{ls}}{\tau_{bs}}\qty(\frac{1}{\tau_{lb}}+\frac{\mu_{j,2}}{\tau_{bs}})\Bigg\}\nonumber\\
    &+\frac{\chi_{a}k_{a\perp1}\chi_{b}k_{b\perp1}}{\omega}\frac{\tau_{ls}}{\tau_{bs}}\qty(\frac{1}{\tau_{la}}+\frac{\mu_{j,1}}{\tau_{as}})+\frac{\chi_{a}k_{a\perp2}\chi_{b}k_{b\perp2}}{\omega}\frac{\tau_{ls}}{\tau_{bs}}\qty(\frac{1}{\tau_{la}}+\frac{\mu_{j,2}}{\tau_{as}})\Bigg]\nonumber
\end{align}
\begin{align}
    &+\int^{\chi_{s}}_{\chi_{l}}d\chi_{b}\int^{\chi_{l}}_{0}d\chi_{a}\int\frac{d\bm{k}_{a}}{(2\pi)^{3}}\int\frac{d\bm{k}_{b}}{(2\pi)^{3}}(-4\omega^{2})\Tilde{\Phi}_{W}(\chi_{a},\bm{k}_{a})\Tilde{\Phi}_{W}(\chi_{b},\bm{k}_{b})e^{ik_{a\parallel}\chi_{a}+ik_{b\parallel}\chi_{b}}
    \nonumber\\
    &\times\exp{i\chi_{a}\bm{k}_{a\perp}\cdot\bm{\theta}_{l,j}+i\chi_{b}\bm{k}_{b\perp}\cdot\frac{\tau_{lb}\bm{\theta}_{l,j}+\tau_{bs}\bm{\theta}_{s}}{\tau_{lb}+\tau_{bs}}}\nonumber\\
    &\times\Bigg[-\sin\Bigg\{\frac{(\chi_{a}\bm{k}_{a\perp})^{2}}{2\omega\tau_{al}}+\frac{(\chi_{b}\bm{k}_{b\perp})^{2}}{2\omega(\tau_{lb}+\tau_{bs})}+\frac{\mu_{j,1}}{2\omega\tau_{ls}}\qty(\chi_{a}k_{a\perp1}+\frac{\tau_{ls}}{\tau_{bs}}\chi_{b}k_{b\perp1})^{2}+\frac{\mu_{j,2}}{2\omega\tau_{ls}}\qty(\chi_{a}k_{a\perp2}+\frac{\tau_{ls}}{\tau_{bs}}\chi_{b}k_{b\perp2})^{2}\Bigg\}\nonumber\\
    &+\sin\Bigg\{\frac{(\chi_{a} k_{a\perp1})^{2}}{2\omega}\qty(\frac{1}{\tau_{al}}+\frac{\mu_{j,1}}{\tau_{ls}})+\frac{(\chi_{a} k_{a\perp2})^{2}}{2\omega}\qty(\frac{1}{\tau_{al}}+\frac{\mu_{j,2}}{\tau_{ls}})\nonumber\\
    &+\frac{(\chi_{b}k_{b\perp1})^{2}}{2\omega}\frac{\tau_{ls}}{\tau_{bs}}\qty(\frac{1}{\tau_{lb}}+\frac{\mu_{j,1}}{\tau_{bs}})+\frac{(\chi_{b}k_{b\perp2})^{2}}{2\omega}\frac{\tau_{ls}}{\tau_{bs}}\qty(\frac{1}{\tau_{lb}}+\frac{\mu_{j,2}}{\tau_{bs}})\Bigg\}\nonumber\\
    &+\frac{\mu_{j,1}\chi_{a}k_{a\perp1}\chi_{b}k_{b\perp1}+\mu_{j,2}\chi_{a}k_{a\perp2}\chi_{b}k_{b\perp2}}{\omega\tau_{bs}}\Bigg].\label{sj2spe}
\end{align}

The Shapiro time delay is defined as $\Delta t_{Wj}\defeq\lim_{\omega\rightarrow\infty}S^{\text{tot}}_{j}(\omega)/\omega$, and can be decomposed as $\Delta t_{Wj}=\Delta t_{Wj}^{(1)}+\Delta t_{Wj}^{(2)}$, where $\Delta t_{Wj}^{(1)}$ and $\Delta t_{Wj}^{(2)}$ represent the first-order and second-order terms in $\Phi$, respectively.
These terms, which are subtracted from $S_{j1}$ and $S_{j2}$ respectively, are given by
\begin{align}
    \Delta t_{Wj}^{(1)}&=-2\int^{\chi_{l}}_{0}d\chi_{a}\int\frac{d\bm{k}_{a}}{(2\pi)^{3}}\Tilde{\Phi}_{W}(\chi_{a},\bm{k}_{a})e^{ik_{a\parallel}\chi_{a}+i\bm{k}_{a\perp}\cdot\chi_{a}\bm{\theta}_{l,j}}\nonumber\\
    &-2\int^{\chi_{s}}_{\chi_{l}}d\chi_{a}\int\frac{d\bm{k}_{a}}{(2\pi)^{3}}\Tilde{\Phi}_{W}(\chi_{a},\bm{k}_{a})e^{ik_{a\parallel}\chi_{a}+i\bm{k}_{a\perp}\cdot\chi_{a}\qty(\frac{\tau_{la}\bm{\theta}_{l,j}+\tau_{as}\bm{\theta}_{s}}{\tau_{la}+\tau_{as}})},\label{shapi1}
    \\
    \Delta t_{Wj}^{(2)}&=4\int^{\chi_{l}}_{0}d\chi_{b}\int^{\chi_{b}}_{0}d\chi_{a}\int\frac{d\bm{k}_{a}}{(2\pi)^{3}}\int\frac{d\bm{k}_{b}}{(2\pi)^{3}}\Tilde{\Phi}_{W}(\chi_{a},\bm{k}_{a})\Tilde{\Phi}_{W}(\chi_{b},\bm{k}_{b})e^{ik_{a\parallel}\chi_{a}+ik_{b\parallel}\chi_{b}+i(\bm{k}_{a\perp}\chi_{a}+\bm{k}_{b\perp}\chi_{b})\cdot\bm{\theta}_{l,j}}\nonumber\\
    &\times \Bigg[\chi_{a}k_{a\perp1}\chi_{b}k_{b\perp1}\qty(\frac{1}{\tau_{bl}}+\frac{\mu_{j,1}}{\tau_{ls}})+\chi_{a}k_{a\perp2}\chi_{b}k_{b\perp2}\qty(\frac{1}{\tau_{bl}}+\frac{\mu_{j,2}}{\tau_{ls}})\Bigg]\nonumber\\
    &+4\int^{\chi_{s}}_{\chi_{l}}d\chi_{b}\int^{\chi_{b}}_{\chi_{l}}d\chi_{a}\int\frac{d\bm{k}_{a}}{(2\pi)^{3}}\int\frac{d\bm{k}_{b}}{(2\pi)^{3}}\Tilde{\Phi}_{W}(\chi_{a},\bm{k}_{a})\Tilde{\Phi}_{W}(\chi_{b},\bm{k}_{b})e^{ik_{a\parallel}\chi_{a}+ik_{b\parallel}\chi_{b}}\nonumber\\
    &\times\exp{i\chi_{a}\bm{k}_{a\perp}\cdot\frac{\tau_{la}\bm{\theta}_{l,j}+\tau_{as}\bm{\theta}_{s}}{\tau_{la}+\tau_{as}}+i\chi_{b}\bm{k}_{b\perp}\cdot\frac{\tau_{lb}\bm{\theta}_{l,j}+\tau_{bs}\bm{\theta}_{s}}{\tau_{lb}+\tau_{bs}}}\nonumber\\
    &\times \Bigg[\chi_{a}k_{a\perp1}\chi_{b}k_{b\perp1}\frac{\tau_{ls}}{\tau_{bs}}\qty(\frac{1}{\tau_{la}}+\frac{\mu_{j,1}}{\tau_{as}})+\chi_{a}k_{a\perp2}\chi_{b}k_{b\perp2}\frac{\tau_{ls}}{\tau_{bs}}\qty(\frac{1}{\tau_{la}}+\frac{\mu_{j,2}}{\tau_{as}})\Bigg]\nonumber\\
    &+4\int^{\chi_{s}}_{\chi_{l}}d\chi_{b}\int^{\chi_{l}}_{0}d\chi_{a}\int\frac{d\bm{k}_{a}}{(2\pi)^{3}}\int\frac{d\bm{k}_{b}}{(2\pi)^{3}}\Tilde{\Phi}_{W}(\chi_{a},\bm{k}_{a})\Tilde{\Phi}_{W}(\chi_{b},\bm{k}_{b})e^{ik_{a\parallel}\chi_{a}+ik_{b\parallel}\chi_{b}}\nonumber\\
    &\times\exp{i\chi_{a}\bm{k}_{a\perp}\cdot\bm{\theta}_{l,j}+i\chi_{b}\bm{k}_{b\perp}\cdot\frac{\tau_{lb}\bm{\theta}_{l,j}+\tau_{bs}\bm{\theta}_{s}}{\tau_{lb}+\tau_{bs}}}\times\Bigg[\frac{\chi_{a}k_{a\perp1}\chi_{b}k_{b\perp1}\mu_{j,1}+\chi_{a}k_{a\perp2}\chi_{b}k_{b\perp2}\mu_{j,2}}{\tau_{bs}}\Bigg].\label{shapi2}
\end{align}

To evaluate the statistical averages of $K_{j2}$ and $S_{j2}$, we encounter terms involving integrals such as $\int^{\chi_{b}}_{0}d\chi_{a}\delta(\chi_{a}-\chi_{b})(\cdots)$, where the integration limit coincides with the peak of the delta function. The treatment of such integrals is summarized as follows.
First, from the normalization of the Dirac delta function, we have:
\begin{align}
    \int^{0}_{-\infty}dx\delta(x)+\int^{\infty}_{0}dx\delta(x)=1.
\end{align}
Since the delta function is an even function $(\delta(-x)=\delta(x))$, the integrals over the negative and positive domains contribute equally, yielding:
\begin{align}
    \int^{0}_{-\infty}dx\delta(x)=\frac{1}{2}.\label{delta}
\end{align}
Next, for an arbitrary smooth function $f(\chi_{a},\chi_{b})$, we consider the integral:
\begin{align}
    I=\int^{\chi_{b}}_{0}d\chi_{a}\delta(\chi_{a}-\chi_{b})f(\chi_{a},\chi_{b}).
\end{align}
By introducing a new variable $\alpha=\chi_{a}-\chi_{b}$, the integral is transformed into:
\begin{align}
    I=\int^{0}_{-\chi_{b}}d\alpha\delta(\alpha)f(\chi_{b}+\alpha,\chi_{b}).
\end{align}
Performing a Taylor expansion of $f(\chi_{b}+\alpha,\chi_{b})$ around $\alpha=0$ gives:
\begin{align}
    I=\int^{0}_{-\chi_{b}}d\alpha\delta(\alpha)\qty[f(\chi_{b},\chi_{b})+\alpha\left.\frac{\partial f}{\partial\chi_{a}}\right|_{\chi_{a}=\chi_{b}}+\cdots].
\end{align}
The first term yields $\frac{1}{2}f(\chi_{b},\chi_{b})$ based on the property derived in \eqref{delta}. The subsequent terms, which involve integrals of the form $\alpha^{n}\delta(\alpha)$ for $n\geq1$, vanish. Consequently, we obtain the following simplification:
\begin{align}
    \int^{\chi_{b}}_{0}d\chi_{a}\delta(\chi_{a}-\chi_{b})f(\chi_{a},\chi_{b})=\frac{1}{2}f(\chi_{b},\chi_{b}).
\end{align}

\section{Derivation of \texorpdfstring{$\langle\abs{F}^{2}\rangle_{W}$}\  \ via Path Integrals} \label{apeC}
By performing the change of variables $\bm{\theta}_{c}=(\bm{\theta}_{1}+\bm{\theta}_{2})/2$ and $\bm{r}=\bm{\theta}_{1}-\bm{\theta}_{2}$ and using the Limber approximation, Eq. \eqref{F2path} can be transformed as follows
\begin{align}
    \langle\abs{F}^{2}\rangle_{W}&=\int\mathscr{D}\bm{\theta}_{c}\mathscr{D}\bm{r}\exp{i\omega\int^{\chi_{s}}_{0}d\chi\chi^{2}\dot{\bm{\theta}}_{c}\cdot\dot{\bm{r}}-i\omega\qty(\psi\qty(\bm{\theta}_{c,l}+\frac{\bm{r}_{l}}{2})-\psi\qty(\bm{\theta}_{c,l}-\frac{\bm{r}_{l}}{2}))}\nonumber\\
    &\times\Bigg\langle\exp{-2i\omega\int^{\chi_{s}}_{0}d\chi(\Phi_{W}(\chi,\bm{\theta}_{s}+\bm{r})-\Phi_{W}(\chi,\bm{\theta}_{s}))}\Bigg\rangle_{W},
\end{align}
where $\bm{\theta}_{c,l}=\bm{\theta}_{c}(\chi_{l})$ and $\bm{r}_{l}=\bm{r}(\chi_{l})$. After decomposing the path integral measure as $\mathscr{D}\bm{\theta}_{c}=d\bm{\theta}_{c,l}\Tilde{\mathscr{D}}\bm{\theta}_{c}$ and $\mathscr{D}\bm{r}=d\bm{r}_{l}\Tilde{\mathscr{D}}\bm{r}$, we perform integration by parts on the kinetic term in the exponent to transfer the derivative, yielding:
\begin{align}
    \langle\abs{F}^{2}\rangle_{W}&=\int d\bm{\theta}_{c,l}d\bm{r}_{l}\int \Tilde{\mathscr{D}}\bm{\theta}_{c}\Tilde{\mathscr{D}}\bm{r}\exp{i\omega\qty(\chi_{s}^{2}\bm{\theta}_{s}\cdot\dot{\bm{r}}_{s}-\chi_{l}^{2}\bm{\theta}_{c,l}\cdot\dot{\bm{r}}_{l+}+\chi_{l}^{2}\bm{\theta}_{c,l}\cdot\dot{\bm{r}}_{l-})}\nonumber\\
    &\times\exp{-i\omega\int^{\chi_{l}}_{0}d\chi\bm{\theta}_{c}\cdot\frac{d}{d\chi}\chi^{2}\frac{d}{d\chi}\bm{r}-i\omega\int^{\chi_{s}}_{\chi_{l}}d\chi\bm{\theta}_{c}\cdot\frac{d}{d\chi}\chi^{2}\frac{d}{d\chi}\bm{r}-i\omega\qty(\psi\qty(\bm{\theta}_{c,l}+\frac{\bm{r}_{l}}{2})-\psi\qty(\bm{\theta}_{c,l}-\frac{\bm{r}_{l}}{2}))}
    \nonumber\\
    &\times\Bigg\langle\exp{-2i\omega\int^{\chi_{s}}_{0}d\chi(\Phi_{W}(\chi,\bm{\theta}_{s}+\bm{r})-\Phi_{W}(\chi,\bm{\theta}_{s}))}\Bigg\rangle_{W}.
\end{align}
Here, quantities with the subscripts $l+$ and $l-$ denote the values evaluated at $\chi_{l}+0$ and $\chi_{l}-0$, respectively.
Since the measure $\Tilde{\mathscr{D}}\bm{\theta}_{c}$ does not include the integration over $\bm{\theta}_{c,l}$, the functional integral can be performed, yielding the following expression that contains a delta functional.
\begin{align}
    \langle\abs{F}^{2}\rangle_{W}&=\int d\bm{\theta}_{c,l}d\bm{r}_{l}\int \Tilde{\mathscr{D}}\bm{r}\exp{i\omega\qty(\chi_{s}^{2}\bm{\theta}_{s}\cdot\dot{\bm{r}}_{s}-\chi_{l}^{2}\bm{\theta}_{c,l}\cdot\dot{\bm{r}}_{l+}+\chi_{l}^{2}\bm{\theta}_{c,l}\cdot\dot{\bm{r}}_{l-})-i\omega\qty(\psi\qty(\bm{\theta}_{c,l}+\frac{\bm{r}_{l}}{2})-\psi\qty(\bm{\theta}_{c,l}-\frac{\bm{r}_{l}}{2}))}\nonumber\\
    &\times\delta^{2}\qty[\omega\frac{d}{d\chi}\chi^{2}\frac{d}{d\chi}\bm{r}(\chi)]
    \Bigg\langle\exp{-2i\omega\int^{\chi_{s}}_{0}d\chi(\Phi_{W}(\chi,\bm{\theta}_{s}+\bm{r})-\Phi_{W}(\chi,\bm{\theta}_{s}))}\Bigg\rangle_{W},\label{C.3}
\end{align}
where the coefficients arising from the path integral measure are absorbed into $\Tilde{\mathscr{D}}\bm{r}$.
Among all possible paths of integration, the delta functional selects only the path for which its argument vanishes. This path satisfies the following differential equation:
\begin{align}
    \frac{d}{d\chi}\chi^{2}\frac{d}{d\chi}\bm{r}(\chi)=0.
\end{align}
The solution to the above equation that satisfies the boundary conditions $\bm{r}(\chi_{l})=\bm{r}_{l}$ and $\bm{r}(\chi_{s})=0$, while remaining finite at $\bm{r}(0)$, can be written as:
\begin{align}
    \bm{r}(\chi) &= 
    \begin{cases}
        \bm{r}_{l} & \text{for } 0 \le \chi < \chi_l \\[8pt]
        \qty(\tau_{ls}/\tau_{ks})\bm{r}_{l} & \text{for } \chi_l \le \chi \le \chi_s
    \end{cases} \label{eq:solution_r}
\end{align}
where $\tau_{ks}=\frac{\chi\chi_{s}}{\chi_{s}-\chi}$. By performing the path integral over $\Tilde{\mathscr{D}}\bm{r}$ and substituting Eq. \eqref{eq:solution_r} into Eq. \eqref{C.3}, we obtain the following expression:
\begin{align}
    \langle\abs{F}^{2}\rangle_{W}&=\mathcal{N}\int d\bm{\theta}_{c,l}d\bm{r}_{l}\exp{i\omega\tau_{ls}(\bm{\theta}_{c,l}-\bm{\theta}_{s})\cdot\bm{r}_{l}-i\omega\qty(\psi\qty(\bm{\theta}_{c,l}+\frac{\bm{r}_{l}}{2})-\psi\qty(\bm{\theta}_{c,l}-\frac{\bm{r}_{l}}{2}))}\nonumber\\
    &\times\Bigg\langle\exp{-2i\omega\int^{\chi_{l}}_{0}d\chi(\Phi_{W}(\chi,\bm{\theta}_{s}+\bm{r}_{l})-\Phi_{W}(\chi,\bm{\theta}_{s}))-2i\omega\int^{\chi_{s}}_{\chi_{l}}d\chi\qty(\Phi_{W}\qty(\chi,\bm{\theta}_{s}+\frac{\tau_{ls}}{\tau_{ks}}\bm{r}_{l})-\Phi_{W}(\chi,\bm{\theta}_{s}))}\Bigg\rangle_{W}.
\end{align}
Here, the coefficients arising from the path integral measure and the delta functional are collectively denoted by $\mathcal{N}$. This normalization constant is determined by the condition that the amplification factor must be unity in the absence of gravitational potentials. In the absence of potentials, $\langle\abs{F}^{2}\rangle_{W}$ can be transformed as
\begin{align}
    \langle\abs{F_{\Phi=0}}^{2}\rangle_{W}&=\mathcal{N}\int d\bm{\theta}_{c,l}d\bm{r}_{l}\exp{i\omega\tau_{ls}(\bm{\theta}_{c,l}-\bm{\theta}_{s})\cdot\bm{r}_{l}}\nonumber\\
    &=\mathcal{N}(2\pi)^{2}\int d\bm{\theta}_{c,l}\delta^{2}(\omega\tau_{ls}(\bm{\theta}_{c,l}-\bm{\theta}_{s}))\nonumber\\
    &=\mathcal{N}\qty(\frac{2\pi}{\omega\tau_{ls}})^{2}.
\end{align}
The condition that this should be unity yields $\mathcal{N}=\qty(\frac{\omega\tau_{ls}}{2\pi})^{2}$. Consequently, $\langle\abs{F}^{2}\rangle_{W}$ can be written as follows
\begin{align}
    \langle\abs{F}^{2}\rangle_{W}&=\qty(\frac{\omega\tau_{ls}}{2\pi})^{2}\int d\bm{\theta}_{c,l}d\bm{r}_{l}\exp{i\omega\tau_{ls}(\bm{\theta}_{c,l}-\bm{\theta}_{s})\cdot\bm{r}_{l}-i\omega\qty(\psi\qty(\bm{\theta}_{c,l}+\frac{\bm{r}_{l}}{2})-\psi\qty(\bm{\theta}_{c,l}-\frac{\bm{r}_{l}}{2}))}\nonumber\\
    &\times\Bigg\langle\exp{-2i\omega\int^{\chi_{l}}_{0}d\chi(\Phi_{W}(\chi,\bm{\theta}_{s}+\bm{r}_{l})-\Phi_{W}(\chi,\bm{\theta}_{s}))-2i\omega\int^{\chi_{s}}_{\chi_{l}}d\chi\qty(\Phi_{W}\qty(\chi,\bm{\theta}_{s}+\frac{\tau_{ls}}{\tau_{ks}}\bm{r}_{l})-\Phi_{W}(\chi,\bm{\theta}_{s}))}\Bigg\rangle_{W}.\label{C.8}
\end{align}
In order to employ the stationary phase approximation for the strong lens, the phase part containing the strong lens potential is defined as
\begin{align}
    f(\bm{\theta}_{c,l},\bm{r}_{l})&\defeq\tau_{ls}(\bm{\theta}_{c,l}-\bm{\theta}_{s})\cdot\bm{r}_{l}-\psi\qty(\bm{\theta}_{c,l}+\frac{\bm{r}_{l}}{2})+\psi\qty(\bm{\theta}_{c,l}-\frac{\bm{r}_{l}}{2}).
\end{align}
The stationary points of this function are found by taking the first derivative of $f$:
\begin{align}
    \bm{\nabla}_{\bm{\theta}_{c,l}}f&=\tau_{ls}\bm{r}_{l}-\bm{\nabla}\psi\qty(\bm{\theta}_{c,l}+\frac{\bm{r}_{l}}{2})+\bm{\nabla}\psi\qty(\bm{\theta}_{c,l}-\frac{\bm{r}_{l}}{2}),\label{C.10}\\
    \bm{\nabla}_{\bm{r}_{l}}f&=\tau_{ls}(\bm{\theta}_{c,l}-\bm{\theta}_{s})-\frac{1}{2}\bm{\nabla}\psi\qty(\bm{\theta}_{c,l}+\frac{\bm{r}_{l}}{2})-\frac{1}{2}\bm{\nabla}\psi\qty(\bm{\theta}_{c,l}-\frac{\bm{r}_{l}}{2})\label{C.11}.
\end{align}
Here, $\bm{\nabla}$ denotes the derivative with respect to the internal variables of the function.
One solution to Eq. \eqref{C.10} is $\bm{r}_{l}=0$, which physically corresponds to considering a single image among the multiple ones produced by the strong lens. Since we have focused on the $j$-th image throughout this paper, we adopt this solution. Substituting this solution into Eq. \eqref{C.11} yields:
\begin{align}
    \tau_{ls}(\bm{\theta}_{l}-\bm{\theta}_{s})-\bm{\nabla}_{\bm{\theta}_{l}}\psi\qty(\bm{\theta}_{l})=0.
\end{align}
Here, we have used the fact that $\bm{r}_{l}=0$ implies $\bm{\theta}_{c,l}=\bm{\theta}_{l}$.
This is precisely the gravitational lens equation, for which we adopt the solution as before. Next, we expand $f(\bm{\theta}_{c,l},\bm{r}_{l})$ up to the second order around the stationary point $(\bm{\theta}_{c,l},\bm{r}_{l})=(\bm{\theta}_{l,j},0)$. Since $f$, its first derivatives, $(\bm{\nabla}_{\bm{\theta}_{c,l}})^{2}f$, and $(\bm{\nabla}_{\bm{r}_{l}})^{2}f$ all vanish at the stationary point, only the following cross terms remain:
\begin{align}
    \frac{\partial^{2}f}{\partial\theta_{c,la}\partial r_{lb}}\Bigg|_{\bm{\theta}_{c,l}=\bm{\theta}_{l,j},\bm{r}_{l}=0}=\tau_{ls}\delta_{ab}-\nabla_{a}\nabla_{b}\psi(\bm{\theta}_{l,j})=\tau_{ls}\qty(\begin{matrix}
        \mu_{j,1}^{-1} & 0\\
        0 & \mu_{j,2}^{-1}
    \end{matrix}).
\end{align}
In the second step, we expressed the result in terms of the magnification $\mu_{j}$ by noting that the second derivative takes the same form as the expansion in Eq. \eqref{spa}. By expanding $f(\bm{\theta}_{c,l},\bm{r}_{l})$ using these relations and $\bm{\theta}_{c,l}=\bm{\theta}_{l,j}+\bm{\alpha}$, and employing delta functions, Eq. \eqref{C.8} can finally be transformed into the following form:
\begin{align}
    \langle\abs{F_{j}}^{2}\rangle_{W}&=\int d\bm{r}_{l}d\bm{\alpha}\qty(\frac{\omega\tau_{ls}}{2\pi})^{2}\exp{i\omega\tau_{ls}(\mu_{j,1}^{-1}r_{l1}\alpha_{1}+\mu_{j,2}^{-1}r_{l2}\alpha_{2})}\nonumber\\
    &\times\Bigg\langle\exp{-2i\omega\int^{\chi_{l}}_{0}d\chi(\Phi_{W}(\chi,\bm{\theta}_{s}+\bm{r}_{l})-\Phi_{W}(\chi,\bm{\theta}_{s}))-2i\omega\int^{\chi_{s}}_{\chi_{l}}d\chi\qty(\Phi_{W}\qty(\chi,\bm{\theta}_{s}+\frac{\tau_{ls}}{\tau_{ks}}\bm{r}_{l})-\Phi_{W}(\chi,\bm{\theta}_{s}))}\Bigg\rangle_{W}\nonumber\\
    &=\int d\bm{r}_{l}\qty(\omega\tau_{ls})^{2}\delta\qty(\omega\tau_{ls}\mu_{j,1}^{-1}r_{l1})\delta\qty(\omega\tau_{ls}\mu_{j,2}^{-1}r_{l2})\nonumber\\
    &\times\Bigg\langle\exp{-2i\omega\int^{\chi_{l}}_{0}d\chi(\Phi_{W}(\chi,\bm{\theta}_{s}+\bm{r}_{l})-\Phi_{W}(\chi,\bm{\theta}_{s}))-2i\omega\int^{\chi_{s}}_{\chi_{l}}d\chi\qty(\Phi_{W}\qty(\chi,\bm{\theta}_{s}+\frac{\tau_{ls}}{\tau_{ks}}\bm{r}_{l})-\Phi_{W}(\chi,\bm{\theta}_{s}))}\Bigg\rangle_{W}\nonumber\\
    &=\abs{\mu_{j}}.
\end{align}
\end{appendices}


\begin{thebibliography}{99}
    \bibitem{LIGO} B. P. Abbott et al., "Observation of Gravitational Waves from a Binary Black Hole Merger" Phys. Rev. Lett. 116, 061102 (2016) [arXiv:1602.03837].
    \bibitem{Nakamura} T. T. Nakamura, S. Deguchi, "Wave Optics in Gravitational Lensing" PTPS.133.137 (1999).
    \bibitem{oguri:wave} M. Oguri, "Strong gravitational lensing of explosive transients" Rept. Prog. Phys. 82, 126901 (2019) [arXiv:1907.06830 [astro-ph.CO]].
    \bibitem{takahashi:wave} R. Takahashi, "Amplitude and Phase Fluctuations for Gravitational Waves Propagating through Inhomogeneous Mass Distribution in the Universe" ApJ 644 80 (2006).
    \bibitem{oguri:probe} M. Oguri and R. Takahashi, "Probing Dark Low-mass Halos and Primordial Black Holes with Frequency-dependent Gravitational Lensing Dispersions of Gravitational Waves" ApJ 901 58 (2020) [arXiv:2007.01936].
    \bibitem{tanaka} S.Tanaka, T.Suyama,"Novel Quantity for Probing Matter Perturbations Below the Fresnel Scale in Gravitational Lensing of Gravitational Waves" Phys. Rev. D 112, 103540 (2025) [arXiv:2503.15066v2].
    \bibitem{mizuno:beyond} M. Mizuno, T. Suyama, "Weak lensing of gravitational waves in wave optics: Beyond the Born approximation" Phys. Rev. D 108, 043511 (2023) [arXiv:2210.02062].
    \bibitem{inamori} M. Inamori, T. Suyama, "Universal relation between the variances of distortions of gravitational waves owing to gravitational lensing" ApJL 918 L30 (2021) [arXiv:2107.02443].
    \bibitem{mizuno:new} M. Mizuno, T. Suyama, and R. Takahashi, "New consistency relations between averages and variances of weakly lensed signals of gravitational waves" Phys. Rev. D 109, 083505 (2024) [arXiv:2309.04114v2].
    \bibitem{oguri:strong} M. Oguri and R. Takahashi, "Amplitude and phase fluctuations of gravitational waves magnified by strong
    gravitational lensing" Phys. Rev. D 106, 043532 (2022) [arXiv:2204.00814v3].
\end{thebibliography}
\end{document}